\documentclass[aps,pre,reprint,nofootinbib,floatfix]{revtex4-2}
\usepackage{amsmath}
\usepackage{amssymb}
\usepackage{graphicx}
\usepackage{mathrsfs}
\usepackage{braket}
\usepackage{svg}
\usepackage[caption=false]{subfig}
\usepackage{hyperref}
\hypersetup{colorlinks,linkcolor=blue,citecolor=blue,urlcolor=blue}

\graphicspath{{figures/}}

\begin{document}


\title{Entropy production from crossing a quantum separatrix}

\author{Peter Stabel}
\author{James R. Anglin}

\affiliation{\mbox{State Research Center OPTIMAS and Fachbereich Physik,} \mbox{RPTU Kaiserslautern-Landau,} \mbox{D-67663 Kaiserslautern, Germany}}

\date{\today}
\begin{abstract}
If the Hamiltonian of a large quantum system depends on a variable parameter, its parameter-dependent energy spectrum may often form a dense lattice of avoided crossings. If the parameter slowly changes, the system's time evolution through the avoided crossings tends in the semi-classical limit to change from adiabatic to diabatic across a narrow energy range that has recently been defined as a \textit{quantum separatrix}. Here we show that in the extreme semi-classical limit of a quasi-continuously dense spectrum, all quantum separatrices form a universal one-parameter family, and that evolution through the separatrix becomes efficiently ergodizing. If the unitary evolution is realistically averaged over small variations in the parameter time dependence, quantum separatrix crossing generates diagonal entropy according to a universal formula. Numerical examples confirm that the results for quasi-continuous spectra remain clearly recognizable in finite systems, with only small finite-size effects.

\end{abstract}

\maketitle

\begin{figure}[t]
    \centering
    \includegraphics[width=\linewidth]{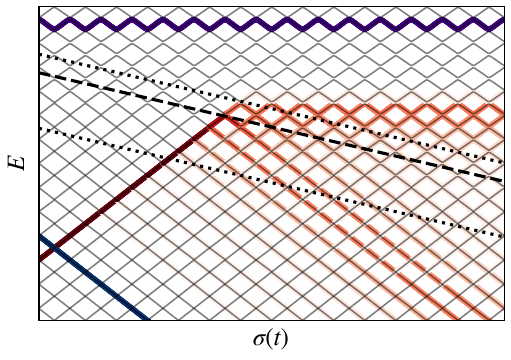}
    \caption{A quantum separatrix. Thin lines indicate discrete energy eigenvalues $E(\sigma)$ of a system whose Hamiltonian depends on a parameter $\sigma$; apparent crossings are actually narrowly avoided due to dynamical tunneling. As $\sigma$ increases with time at a given slow rate, energy eigenstates in one portion of the $(\sigma,E)$ plane evolve adiabatically through each avoided crossing, like the state whose probability is indicated in purple. In another $(\sigma,E)$ region, states instead evolve diabatically, as illustrated in blue. The transition between these two behaviors at avoided level crossings occurs across a quantum separatrix zone, within which probability forks at each crossing, as shown in red. The black dashed line marks the quantum separatrix contour where $P = 1/e$. The dotted lines indicate the (arbitrary) low and high probability thresholds, $P_{\mathrm{low}}  = 0.01$ and $P_\mathrm{high}=0.99$. We call the region of avoided crossings that fall within these thresholds the quantum separatrix zone. }
    \label{fig:qsep_illustration}
\end{figure}

\section{Introduction}
Evolution with multiple time scales is an unavoidable complexity in many important physical systems, which fortunately is simplified by adiabaticity.
If a Hamiltonian depends on a slowly time-dependent parameter, the adiabatic theorem ensures that the system's energy changes in one-to-one dependence on the parameter, so that knowing only the system's energy at one time allows us to predict the energy at later times, at least up to small post-adiabatic corrections, even if we do not know anything about the system's state apart from its energy. In particular, if the slowly changing parameter later returns to its initial value, the system's energy will also return to its initial value. 

If the slow parameter change brings the system close to a separatrix, however, then the breakdown of adiabaticity at the separatrix disrupts this reversible behavior, by changing the system's energy in a way that depends sensitively on rapidly evolving coordinates. If these rapid coordinates cannot be precisely observed or controlled, the energy change at the separatrix can be effectively probabilistic, and an ensemble of trajectories that is averaged over fast motion can increase in coarse-grained entropy, even though the microscopic evolution is Hamiltonian. In the limit of infinitely slow parameter change, the classical probabilities of different energies, and the effective entropy increase from separatrix crossing, can be derived from Liouville's theorem; if parameter change is sufficiently but only finitely slow, the corrections are small \cite{NEISHTADT1975594,Neishtadt_2017,henrard_capture_1982, LiouvilleControl}. These effects have recently been shown to imply \textit{probabilistic hysteresis} as a microscopic form of irreversibility \cite{burkle_how_2019,buerkle_trimer}.

Any form or sense of irreversibility that is fully microscopic must be quantum mechanical, however, and extending our classical understanding of post-adiabatic separatrix crossing into quantum mechanics is complicated by the fact that the classical and adiabatic limits do not commute \cite{Berry_1984}. Finitely slow forward-and-back parameter sweeps in a quantum system may quite closely approach the probabilities for infinitely slow sweeps through a classical separatrix, but for even slower sweeps the quantum evolution through the corresponding energy range instead becomes steadily more adiabatic, with vanishing entropy growth \cite{buerkle_qdimer1, buerkle_qdimer2}.

The ambiguity of classical and adiabatic limits for separatrices has recently been resolved by identifying the \textit{quantum separatrix} as a narrow region in energy- and parameter-space whose location and shape depend not only on the parameter-dependent Hamiltonian, but also on the rate at which the parameter changes in time \cite{Stabel_2022,qknh2,harpermodel}. Here we will show that in the semi-classical limit of a quasi-continuously dense energy level spectrum, evolution through any quantum separatrix takes a universal form depending on a single real parameter that characterizes the separatrix, and which determines whether and how entropy changes as the separatrix is crossed.  

A typical quantum separatrix is illustrated in Fig.~\ref{fig:qsep_illustration}. The spectrum consists almost everywhere of two overlapping ladders of energy eigenvalues $E^{\mathcal{A}}_m$ and $E^{\mathcal{B}}_n$, which depend differently on the time-dependent parameter $\sigma$, and therefore form a lattice in the $E-\sigma$ plane, with many apparent crossings of $E^{\mathcal{A}}_m$ and $E^{\mathcal{B}}_n$. The two ladders of $\mathcal{A}$ and $\mathcal{B}$ states could represent states localized in each side of a double well with $\sigma$-dependent asymmetry, but they could also arise in other ways, for example as molecular states in which two different electronic states are both dressed by vibrational excitations. 

The crossings are actually narrowly avoided, with small minimum energy differences that are typically due to dynamical tunneling. For brevity we will often refer to the avoided crossings simply as crossings, and we will use the diabatic basis of $E^{\mathcal{A}}_m$ and $E^{\mathcal{B}}_n$ states that may be interpolated linearly through each crossing, even though near each crossing the exact adiabatic eigenstates are superpositions of the two nearly degenerate diabatic states. If $\sigma(t)$ changes slowly in time compared to the typical time scale defined by the spacing between levels, then evolution is adiabatic, except possibly near the lattice's crossings. Since adiabaticity thus breaks down only in separate avoided crossings of pairs of levels, the standard Landau-Zener post-adiabatic theory can be applied to evolve accurately through each of the avoided crossings.

\subsection{Incoherent average evolution}
Evolution through an avoided crossing is in principle a coherent process, and the Landau-Zener formula is for the amplitudes in a coherent superposition of diabatic and adiabatic evolutions, rather than a probability of one or the other.
A further feature of the multiple-scale evolution, however, is that over the comparatively long adiabatic intervals between avoided crossings, the amplitudes for superposed instantaneous energy eigenstates will accumulate large relative phases. Proportionally very small perturbations to either the Hamiltonian or the parameter change rate can easily change these large between-crossing phases by at least order $2\pi$. In any long sequence of repeated experiments or large ensemble of similar systems, therefore, the phases of different energy states in different runs or individual systems will in realistic practice be independently random at every avoided crossing. In adiabatic evolution through a lattice of avoided level crossings, therefore, the average density operator, which represents all properties which are averaged over many runs or system examples, is in practice bound to be a probabilistic mixture of instantaneous energy eigenstates, with time-dependent probabilities $P^\mathcal{A}_m$ and $P^\mathcal{B}_n$ for the $E^\mathcal{A}_m$ and $E^\mathcal{B}_n$ states, respectively, rather than a pure quantum state. 

Evolving in time with slow $\sigma(t)$, furthermore, these average probability distributions will be constant, except for probabilistic splitting and mixing at each avoided crossing. In particular, over the comparatively short time in which the change in $\sigma$ brings two diabatic levels $E^\mathcal{A}_m$ and $E^\mathcal{B}_n$ through their crossing $(m,n)$, the two probabilities will change according to
\begin{equation}\label{LZ1}
    \left(\begin{matrix}P^\mathcal{A}_m\\ P^\mathcal{B}_n\end{matrix}\right)\longrightarrow  \left( \begin{matrix}P_{mn} & 1-P_{mn}\\ 1-P_{mn} & P_{mn}\end{matrix}\right)\left(\begin{matrix}P^\mathcal{A}_m\\ P^\mathcal{B}_n\end{matrix}\right)\;,
\end{equation}
where $P_{mn}$ is the Landau-Zener probability for diabatic evolution through the avoided crossing $(m,n)$.

Perturbations large enough to induce $\mathcal{O}(2\pi)$ phases between crossings can easily be too small to significantly affect the Landau-Zener branching probabilities during a crossing; it is this regime of random between-crossing phases, but a consistent set of $P_{mn}$, that we address in this article. The scenario is by no means exotic, but generic in a wide range of physical systems with multiple time scales. Our contribution is to show that it has a simple semi-classical limit.
\begin{figure*}
    \centering
    \includegraphics[width=\linewidth]{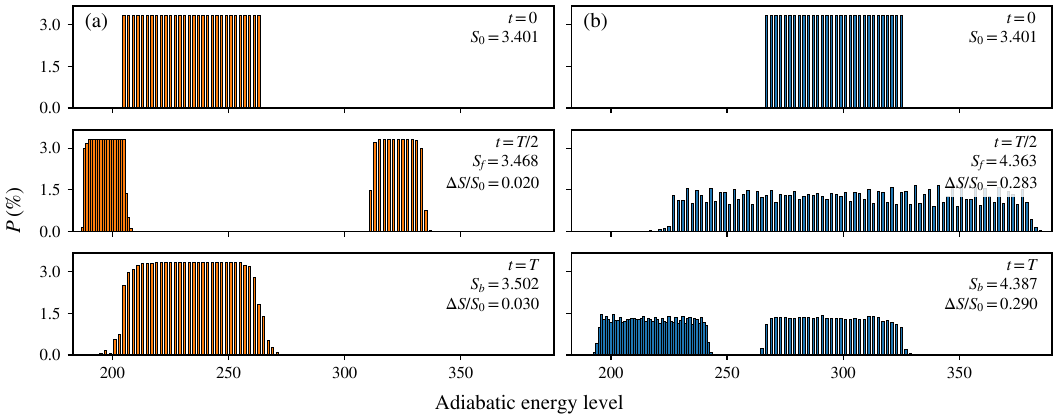}
    \caption{Probability distributions over the instantaneous eigenstates of a time-dependent quantum pendulum for two forward-backward sweep protocols. Protocol 1 ((a), orange, $\lambda\approx0.44$ for the forward sweep): the area enclosed by the separatrix in the corresponding classical model grows during the forward sweep and shrinks during the backward sweep. In protocol 2 ((b), blue, $\lambda\approx -0.44$ for the forward sweep) the order is reversed.  Each initial ensemble contains 30 equally populated states. Protocol 1 approximately preserves the number of appreciably occupied states, whereas protocol 2 spreads the probability over substantially more states. Thus the diagonal entropy increase after the forward sweep is very small under protocol 1 but substantial under protocol 2. Fig.~\ref{fig:S_diag} in Appendix~\ref{app:pendulum} shows the corresponding evolution of diagonal entropy over time under the two protocols.}
    \label{fig:numerical_example}
\end{figure*}
\subsection{The semi-classical limit}
In \cite{Stabel_2022} and \cite{qknh2} we have shown that in the semi-classical limit the level lattice in the $E-\sigma$ plane becomes dense in classical terms, in the sense that there are $\mathcal{O}(\hbar^{-1})$ levels within a given narrow energy range, and also locally regular, with the unit cell only changing significantly over $\mathcal{O}(\hbar^{-1})$ lattice intervals. Furthermore, if the avoided crossings in the lattice are labeled with discrete coordinates $(m,n)$ and an arbitrary point in the lattice is defined as the origin $m=n=0$, then at least in cases where the crossings are avoided due to dynamical tunneling, the Landau-Zener probability for diabatic evolution through a given crossing takes the double-exponential form
\begin{equation}\label{Pmn}
    P_{mn} = \exp\left({-Z e^{mX+nY}}\right),
\end{equation}
for $Z>0$ and real parameters $X$, and $Y$ that are derivable from the classical Hamiltonian and from the rate of change of $\sigma(t)$. 
To leading semiclassical order, the parameters are \cite{Stabel_2022,qknh2}
\begin{align}
X&=-2\pi\frac{[\mathcal K,\tilde S_B]_{E,\sigma}}{[\tilde S_A,\tilde S_B]_{E,\sigma}},
\quad
Y=-2\pi\frac{[\mathcal K,\tilde S_A]_{E,\sigma}}{[\tilde S_B,\tilde S_A]_{E,\sigma}},
\\
Z&=\frac{\pi\hbar\,e^{-2\mathcal K/\hbar}}
{2|\dot\sigma[\tilde S_A,\tilde S_B]_{E,\sigma}|},
\end{align}
where $[F,G]_{E,\sigma}=\partial_EF\,\partial_\sigma G-\partial_\sigma F\,\partial_EG$, and all quantities are evaluated at a reference crossing $(E_{00},\sigma_{00})$. Here $\tilde S_{A,B}/\pi$ are the phase-corrected action variables of the two classical orbit families, while $\mathcal K$ is the integral governing barrier tunneling or above-barrier reflection. The double exponential follows by inserting the exponentially small semiclassical gap into the Landau–Zener formula and expanding the actions locally over the crossing lattice.
While $X,Y,Z$ actually depend on $m,n$, in the semi-classical limit they only change significantly if $m$ and $n$ change by $\mathcal{O}(\hbar^{-1})$, and so a semi-classical lattice of avoided crossings is fully characterized locally by its parameters $X,Y,Z$.

The double-exponential form (\ref{Pmn}) of $P_{mn}$, and the density of the spectrum in the semi-classical limit, ensure that evolution through the lattice of avoided crossings changes from diabatic ($P_{mn}\sim 1$) to adiabatic ($P_{mn}\sim 0$) within a narrow energy range with width of order $\hbar$. This narrow region in the $(E,\sigma)$ plane is the quantum separatrix, and (\ref{Pmn}) has been used in \cite{Stabel_2022} to derive the quantum analog of the Kruskal-Neishtadt-Henrard (KNH) theorem \cite{NEISHTADT1975594,Neishtadt_2017,henrard_capture_1982, LiouvilleControl} which governs the proportions into which an ensemble may split as it crosses a separatrix.

With average adiabatic evolution being a succession of probabilistic mixings between pairs of state probabilities, the possibility of increase in diagonal entropy \cite{POLKOVNIKOV2011486,quench,correlations,quench_magnetism,qbit}
\begin{equation}
    S_d = - \Big(\sum_m P^\mathcal{A}_m \ln P^\mathcal{A}_m + \sum_n P^\mathcal{B}_n \ln P^\mathcal{B}_n\Big)
\end{equation}
is thus clear whenever $P_{mn}$ are between zero and one, as they are within a quantum separatrix zone. Exactly how and even whether $S_d$ actually increases, however, is non-trivial. 
For example, it could be that both $X$ and $Y$ in $P_{mn}$ are large while $Z$ is $\mathcal{O}(1)$, so that $P_{mn}$ changes abruptly, between one $(m,n)$ and its neighbor, from essentially zero to essentially one. In this limit, $S_d$ remains constant because although the $P^\mathcal{A}_m$ and $P^\mathcal{B}_n$ may be redistributed among instantaneous energies $E^{\mathcal{A}}_m$ and $E^{\mathcal{B}}_n$, the same set of probabilities is retained. An initial ensemble can be split up into separate pieces, like a block of cheese being grated, but it is not stretched or compressed. 
\subsection{Example of reversible splitting vs irreversible spreading of the initial distribution}
On the other hand, four examples of evolution through a quantum separatrix with $X$ and $Y$ both of $\mathcal{O}(1)$ are illustrated in Fig.~\ref{fig:numerical_example} (by accounting for the forward-and backward parameter sweep). The numerical evolution shown does not simply assume our incoherently adiabatic scenario, but rather coherently evolves an ensemble of realizations of the quantum pendulum (see also \cite{qknh2}) with slightly different rates of slow parameter change (see Appendix \ref{app:pendulum}); ensemble-averaged probabilities are then shown.   Two different initially uniform finite ensembles of states of the same system both undergo a slow parameter sweep from time $t=0$ to $t=T/2$ that brings the ensembles through a separatrix; the parameter sweep is then reversed, to bring the two ensembles back through the separatrix in reverse, and return the slowly time-dependent Hamiltonian parameter to its initial value, $\sigma(T)=\sigma(0)$. 

Under protocol 1 (orange in Fig. \ref{fig:numerical_example}) the ensemble has split into two distinct peaks at the intermediate reverse point $t=T/2$, which are merged again during the reverse sweep and the final probability distribution at $t=T$ is close to the initial one except it is slightly broadened. Under protocol 2 (blue in Fig. \ref{fig:numerical_example}), the probability distribution does not split during the forward sweep, but during the reverse sweep, resulting in a final distribution that shows two distinct peaks, so the final probability distribution is significantly different from the initial distribution. In this sense, the evolution under protocol 1 can be viewed as nearly reversible while the evolution under protocol 2 shows \textit{probabilistic hysteresis} and is thus irreversible. 
At the sweep reversal point $t=T/2$, and then even more at the final time $t=T$, the probability distributions have become slightly non-uniform everywhere, and noticeably spread at their edges. No precise rule for the exact entropy change is apparent in any case. In the intermediate state of ensemble 2 (blue), however, a significant overall reduction and spreading of probability stands out above the small variations and edge effects, with a correspondingly substantial increase in diagonal entropy after the forward crossing of the separatrix. Why does this larger kind of entropy increase not occur in ensemble 1 (orange), or in the reverse crossing in ensemble 2? And what determines just how large is the larger share of the entropy increase in ensemble 2, beyond the smaller effects? 
We answer these questions in a similar way to explaining a sharp crossover in a finite system by relating it to a phase transition in an infinite system: we analyze an extreme limit. We consider the limit in which $X$ and $Y$ are both so small that $P_{mn}$ changes from 0 to 1 very gradually, over a large range of $m,n$, while simultaneously assuming that our Landau-Zener lattice is regular enough that $X$ and $Y$ remain effectively constant over this whole range. This extreme limit may only rarely be strictly valid, but the behavior that we can strictly derive in this limit will also appear clearly in less extreme cases as a large effect with smaller corrections, as in Fig.~\ref{fig:numerical_example}.

\section{Continuum probability transport}
\subsection{Derivation of an effective differential equation}
Within our adiabatic scenario, time evolution is fully described by determining $P^\mathcal{A}_m$ and $P^\mathcal{B}_n$ just \emph{before} each crossing $(m,n)$, since the probabilities only change at the crossings, and each crossing occurs at a specific value of $\sigma=\sigma_{mn}$, which we assume to be a monotonic slow function of time. We therefore map the time-dependent probabilities onto functions $A_{mn},B_{mn}$ defined on the $m,n$ grid of crossings:
\begin{equation}
    P^\mathcal{A}_m(t)\longrightarrow A_{mn}\qquad P^\mathcal{B}_n(t)\longrightarrow B_{mn}\;.
\end{equation}

In this notation, Eq.~(\ref{LZ1}) for the Landau-Zener transition at crossing $(m,n)$ becomes
\begin{equation}
\label{transition_mn}\left(\begin{matrix}A_{m,n+1}\\ B_{m-1,n}\end{matrix}\right)=  \left( \begin{matrix}P_{mn} & 1-P_{mn}\\ 1-P_{mn} & P_{mn}\end{matrix}\right)\left(\begin{matrix}A_{mn}\\ B_{mn}\end{matrix}\right)\;.
\end{equation}

We then consider cases in which an initial ensemble of many adjacent states has $P^\mathcal{A}_m$ and/or $P^\mathcal{B}_n$ that vary only slowly with $m$ or $n$. The whole ensemble is initially outside the separatrix zone, enters the zone as $\sigma(t)$ slowly increases, and may then eventually exit the zone. Since the $P_{mn}$ are initially either very close to 0 or 1, and change slightly from one crossing to the next, the grid functions $A_{mn},B_{mn}$ will be slowly varying functions of both $m$ and $n$.

Finally we consider the limit where the local Landau-Zener lattice parameters $X$ and $Y$ are both small:
\begin{equation}
    X = (1-\lambda)\varepsilon\qquad Y = (1+\lambda)\varepsilon
\end{equation}
for $\varepsilon\ll 1$, which defines
\begin{equation}
    \varepsilon = \frac{X+Y}{2}\qquad \lambda=\frac{Y-X}{X+Y}.
\end{equation}
We choose the lattice orientation such that $\varepsilon>0$. The parameter $\lambda$ will turn out to be the only characteristic of a semi-classical separatrix. We then interpret the slowly varying $A_{mn},B_{mn}$ as the values at a discrete grid of points of functions defined in a continuous plane:
\begin{align}
    [A_{mn},B_{mn}] &=[A(\tau,\eta),B(\tau,\eta)]\Big|_{\tau\to \tau_{mn},\eta\to\eta_{mn}}\nonumber\\\label{eta}
    \eta_{mn} &= \varepsilon(m+n)\\\label{tau}
    \tau_{mn} &=\varepsilon (n-m)\;,
\end{align}
The continuous functions $A(\tau,\eta)$ and $B(\tau,\eta)$ interpolate occupation probabilities of individual states. For constant $\tau$, neighboring states in either diabatic sector are separated by $2\varepsilon$ in $\eta$, \textit{i.e.} $\Delta\eta = \varepsilon(\Delta m + \Delta n)= 2\varepsilon\Delta m$ (since at fixed $\tau$, $\Delta m = \Delta n$). Consequently, sums over states become integrals with measure $d\eta/(2\varepsilon)$, and the normalization becomes $\frac{1}{2\varepsilon}\int d\eta\,
[A(\tau,\eta)+B(\tau,\eta)]=1$. The corresponding probability densities with respect to $\eta$ are then $A/(2\varepsilon)$ and $B/(2\varepsilon)$.
The left hand side of \eqref{transition_mn} in the continuous coordinates becomes
\begin{equation}\label{eq:A_B_prime}\left(\begin{matrix}A_{m,n+1}\\B_{m-1,n}\end{matrix}\right) = \left(\begin{matrix}A(\tau_{mn}+\varepsilon,\eta_{mn}+\varepsilon)\\B(\tau_{mn}+\varepsilon,\eta_{mn}-\varepsilon)\end{matrix}\right)\;.\end{equation}
Taylor expansion up to first order in $\varepsilon$ yields a partial differential equation in $(\eta,\tau)$: 
\begin{equation}
  \left(\begin{matrix}  \frac{\partial}{\partial\tau}+  \frac{\partial}{\partial\eta} & 0\\0 &   \frac{\partial}{\partial\tau} -   \frac{\partial}{\partial\eta} \end{matrix}\right)\left(\begin{matrix}A\\B\end{matrix}\right) = -\frac{1-P}{\varepsilon}\left(\begin{matrix}1 & -1\\-1 & 1\end{matrix}\right)\left(\begin{matrix}A\\B\end{matrix}\right)\;.
\end{equation}
The coordinate $\tau$ serves as a dimensionless evolution coordinate
on the crossing lattice. Indeed, at fixed $\eta$ the crossing parameter
$\sigma$ varies linearly with $\tau$, as shown in
Appendix~\ref{app:lattice}; hence a monotonic sweep of $\sigma$ gives
$\tau$ a time-like orientation. A $\sigma$-dependent c-number shift of
the Hamiltonian can be chosen so that the shifted energy at each
crossing is proportional to $\eta$. For unequal diabatic level
spacings, constant-$\tau$ lines do not in general coincide with constant-$\sigma$
lines.

The argument of the transition probability according to Eq. \eqref{Pmn} only depends on $\ln(Z)+mX+nY$. We define the shifted coordinate 
 \begin{align}
    s =mX+nY + \ln(Z/\varepsilon)= \eta + \lambda\tau + \ln(Z/\varepsilon)
\end{align} 
which allows us to write the transition probability as $P(s)=\exp(-\varepsilon e^s)$. At fixed s, the coefficient $\varepsilon^{-1}(1-P(s))$ satisfies
\begin{equation}
    \frac{1-P(s)}{\varepsilon}=e^s + \mathcal{O}\left(\varepsilon e^{2s}\right).
\end{equation} 
The coefficient $\varepsilon^{-1}(1-P)$ is of order one at $s=0$ and becomes large for $e^s\gg1$. It remains large even when $\varepsilon e^s\gtrsim1$, where the Taylor expansion ceases to be accurate because in that regime the exact coefficient is of order $\varepsilon^{-1}$. This is the adiabatic regime associated with the simple asymptotic solution we later present in Eq. \eqref{above_sep}. For $1\ll e^s\ll\varepsilon^{-1}$, the coefficient is already large while the relative error in replacing $(1-P)/\varepsilon$ by $e^s$ is still small, of order $\varepsilon e^s$. Keeping the leading term yields
\begin{align}\label{eq:universaleq}
&\Bigg[(\lambda^2-1)\frac{\partial}{\partial s} +\left(\begin{matrix}\lambda-1 & 0\\0 & \lambda+1\end{matrix}\right)\frac{\partial}{\partial \tau}\Bigg]\left(\begin{matrix}A\\B\end{matrix}\right)\\\nonumber
&= -e^s\left(\begin{matrix}\lambda-1 & 1-\lambda\\-\lambda-1 & \lambda+1\end{matrix}\right)\left(\begin{matrix}A\\ B\end{matrix}\right)\;.
\end{align}
In our extreme semi-classical limit, the only way in which Eq.~(\ref{eq:universaleq}) can differ for different systems is through the single dimensionless real parameter $\lambda$. Since (\ref{eq:universaleq}) contains no explicit $\tau$-dependence, it can be simplified further by expressing $A$ and $B$ in terms of the Fourier ansatz
\begin{equation}
\begin{pmatrix}
\label{separation}
A(\tau,s)\\
B(\tau,s)
\end{pmatrix}\longrightarrow e^{-i\nu\tau}
\begin{pmatrix}
a_\nu(s)\\
b_\nu(s)
\end{pmatrix}\;.
\end{equation}
This reduces (\ref{eq:universaleq}) to a two-component linear ODE of first order, which can readily be solved numerically, and for which the asymptotic behavior at large and small $\nu$ and $s$ can be derived analytically by standard methods:
\begin{align}
    \frac{d}{ds}\begin{pmatrix}
        a_\nu(s)\\b_\nu(s)
    \end{pmatrix} = (\lambda^2-1)^{-1}\Bigg[i\nu\begin{pmatrix}
        \lambda-1&0\\0&\lambda+1
    \end{pmatrix}\nonumber\\
    -e^s\begin{pmatrix}
        \lambda-1&1-\lambda\\-(\lambda+1)&\lambda+1
    \end{pmatrix}\Bigg]\begin{pmatrix}
        a_\nu(s)\\b_\nu(s)
    \end{pmatrix}
    \label{eq:effective_ODE}
\end{align}
and our analysis excludes the edge cases $\lambda =0, \pm1$. This final equation for incoherent adiabatic evolution of an ensemble through a quantum separatrix even somewhat resembles the original Landau-Zener problem of Schr\"odinger evolution through an avoided crossing, except that the evolution now is in the skewed energy-time coordinate $s$ rather than pure time, involves Fourier components of probabilities rather than amplitudes, and is non-Hermitian.

We distinguish three asymptotic dynamical sectors according to how states propagate through the crossing lattice. On the diabatic side of the quantum separatrix, a state follows either the A ladder at fixed $m$ or the B ladder at fixed $n$; these states form sectors $\mathcal A$ and $\mathcal B$, respectively. On the adiabatic side, a state follows an instantaneous eigenvalue branch through successive avoided crossings, changing between the two diabatic ladders at every avoided crossing. These adiabatic branches form sector $\mathcal C$. The sectors therefore describe different asymptotic parts of the same crossing network. In the following “above” and “below” the quantum separatrix refer to the adiabatic and diabatic regions outside of the quantum separatrix zone, respectively, as illustrated in Fig.~\ref{fig:qsep_illustration}. Since $P(s)$ decreases with increasing $s$, the adiabatic side corresponds to larger $s$. In a double-well model, increasing energy strengthens tunneling and decreases the diabatic transition probability, so the adiabatic side lies at higher energies. In the pendulum example we show in Fig.~\ref{fig:numerical_example}, increasing energy instead weakens above-barrier reflection and increases the diabatic transition probability, reversing this energy ordering.

\subsection{Asymptotic solutions above the quantum separatrix}\label{above_sep_asymptotic}
The right hand side of Eq.~(\ref{eq:effective_ODE}) has the two eigenvalues $\mu_\pm = (\lambda^2-1)^{-1}[\lambda(i\nu-e^s)\pm\sqrt{\lambda^2e^{2s}-2i\nu e^s-\nu^2}]$. Thus at large positive s, the equation is solved in terms of two eigenmodes, one of which has an eigenvalue with the large leading real part $-2e^s\frac{\lambda}{\lambda^2-1}$ and eigenvector that approaches $(1-\lambda,\,1+\lambda)^T$. The second large-$s$ asymptotic eigenvalue is $i\nu/\lambda+\mathcal{O}(\nu^2e^{-s})$ with the asymptotic eigenvector $(1,1)^T$. Since the first mode either decays rapidly or diverges exponentially with increasing $s$, only the latter remains as a physical asymptotic solution whenever the growing mode is excluded. Hence the asymptotic mode solution is of the form 
\begin{equation}
\label{above_sep}
    \left(\begin{matrix}
A(\tau,s)\\ B(\tau,s)\end{matrix}\right)\propto \left(\begin{matrix}
1\\ 1\end{matrix}\right)e^{i\nu(\frac{s}{\lambda}-\tau)}
\end{equation}
Thus the only surviving asymptotic mode has $A=B=A\!\left(\tau-\frac{s}{\lambda}\right)=
A\!\left(-\frac{\eta+s_0}{\lambda}\right)$ with $s_0=\ln(Z/\varepsilon)$. For fixed local lattice parameters $X,Y,Z$, its dependence in the crossing indices is only through $m+n$. In this regime, adiabatic following brings a state through successive avoided crossings by alternating between diabatic ladders $A$ and $B$. Neither $m$ nor $n$ remains fixed through such an adiabatic path. Hence, we label consecutive adiabatic states by an integer $j$ which remains fixed along each branch.

For $\lambda>0$ this mode propagates away from the quantum separatrix and therefore represents a transmitted wave, whereas for $\lambda<0$ it propagates towards the separatrix and constitutes an incident wave.
\subsection{Asymptotic solutions below the quantum separatrix}
At large negative $s$ the asymptotic solutions of \eqref{eq:effective_ODE} are given by
\begin{equation}
    \left(\begin{matrix}
A(\tau,s)\\B(\tau, s)\end{matrix}\right) = e^{-i\nu\tau}\left(\begin{matrix}
\alpha_\nu e^{i\frac{\nu}{\lambda+1}s}\\ \beta_\nu e^{i\frac{\nu}{\lambda-1}s}\end{matrix}\right)
\end{equation}
Transforming back to the original variables yields $A(\tau_{mn},\eta_{mn})=A(m),B(\tau_{mn},\eta_{mn})=B(n)$. $A$ and $B$ decouple from each other and reproduce the diabatic transition regime in which $P^\mathcal{A}_m$ and $P^\mathcal{B}_n$ are independently conserved. 

For $\lambda<-1$, $A$ and $B$ both propagate only in the negative $s$ direction, and hence are not incident waves approaching the separatrix. For $-1<\lambda<1$, there can be an incident wave $(a(s),b(s))\propto (1,0)$, but the $b$-component can only be an outgoing wave, transmitted or reflected, while for $\lambda>1$ there are two independent incident waves $a$ and $b$.

The range of small $|s+\ln\varepsilon|$, a finite-width strip running at some angle across the $(m,n)$ plane, is the quantum separatrix zone. The transmission problem of connecting the asymptotic solutions to (\ref{eq:universaleq}) as $s\to-\infty$ and $s\to +\infty$ determines how ensembles cross the quantum separatrix: it determines whether they are split, and whether or not they are stretched, increasing entropy. All these questions are determined by $\lambda$. As the only parameter of the quantum separatrix, $\lambda$ defines the slope of the quantum separatrix in the $(m,n)$ plane, and thereby determines geometrically which ensembles will enter and cross the separatrix, and which simply propagate away from it, like the state shown in blue in Fig.~\ref{fig:qsep_illustration}.
\section{Ensemble splitting and geometry-dependent diagonal entropy growth}
\begin{table*}
 \centering
  \small
 \setlength{\tabcolsep}{5pt}
 \renewcommand{\arraystretch}{1.8}
 \begin{tabular}{ccl}
 \toprule
 & Incident & Outgoing\\
 \hline
 $\lambda<-1$ & $P_C^{\rm in}(j)= f_C(\varepsilon j)$ &
 $\begin{aligned}
 P_A^{\rm out}(m)&=
 f_C\!\left(\frac{\lambda}{\lambda+1}2\varepsilon m\right),\\
 P_B^{\rm out}(n)&=
 f_C\!\left(\frac{\lambda}{\lambda-1}2\varepsilon n\right)
 \end{aligned}$ \\
 $-1<\lambda<0$ & $\begin{aligned}P_A^{\rm in}(m)&=f_A(2\varepsilon m),\\ P_C^{\rm in}(j)&= f_C(\varepsilon j)\end{aligned}$ &
 $\begin{aligned}
 P_B^{\rm out}(n)&=
 \frac{1+\lambda}{1-\lambda}
 f_A\!\left(\frac{1+\lambda}{\lambda-1}2\varepsilon n\right)+\frac{2\lambda}{\lambda-1}
 f_C\!\left(\frac{\lambda}{\lambda-1}2\varepsilon n\right)
 \end{aligned}$ \\
 $0<\lambda<1$ & $P_A^{\rm in}(m)=f_A(2\varepsilon m)$ &
 $\begin{aligned}
 P_B^{\rm out}(n)&=
 f_A\!\left(\frac{\lambda+1}{\lambda-1}2\varepsilon n\right),\\
 P_C^{\rm out}(j)&=
 f_A\!\left(\frac{\lambda+1}{\lambda}\varepsilon j\right)
 \end{aligned}$ \\
 $\lambda>1$ & $\begin{aligned}P_A^{\rm in}(m)&=f_A(2\varepsilon m),\\ P_B^{\rm in}(n)&=f_B(2\varepsilon n)\end{aligned}$ &
 $\begin{aligned}
 P_C^{\rm out}(j)&=
 \frac{\lambda+1}{2\lambda}
 f_A\!\left(\frac{\lambda+1}{\lambda}\varepsilon j\right)+\frac{\lambda-1}{2\lambda}
 f_B\!\left(\frac{\lambda-1}{\lambda}\varepsilon j\right)
 \end{aligned}$ \\
 \hline
 \end{tabular}
  \caption{Map of incoming probability profiles to outgoing probability profiles for each regime of $\lambda$. Here, a constant coordinate offset by $s_0 = \ln(Z/\varepsilon)$ has been omitted. To evaluate the maps at the original state indices, replace $2\varepsilon m$, $2\varepsilon n$, $\varepsilon j$ inside every profile argument by $2\varepsilon m+s_0$, $2\varepsilon n+s_0$, $\varepsilon j+s_0$ before applying the scaling in the arguments. Restoring these offsets does not change the integrated sector probabilities or single-input entropy formulas. They must, however, be retained when evaluating the overlap of two incident profiles.}
  \label{tab:explicit_profile_maps}
\end{table*}
The asymptotic solutions to the continuum scattering problem \eqref{scattering_eta} together with the leading order transfer coefficients \eqref{firstorder_TR} that we derive in Appendix \ref{sec:matched_asymptotics} determine how incoming probability profiles are redistributed between the outgoing sectors. We present these maps for each case of $\lambda$. We use them to distinguish cases where the incoming probability splits between two sectors and cases where two incoming probability sectors merge into a single outgoing sector. We give a formula for the increase in diagonal entropy in the case where we restrict to one incoming sector in the merging scenario.  
\subsection{Probability profiles and total sector probabilities}
We denote the occupation probability of individual diabatic states in sectors $\mathcal{A}, \mathcal{B}$ by $P_A(m), P_B(n)$ and use the relation to the continuum coordinates $\eta-\tau = 2\varepsilon m$ and $\eta+\tau = 2\varepsilon n$. As we discussed in Sec.~\ref{above_sep_asymptotic}, the asymptotic solution far above the quantum separatrix in the adiabatic sector $\mathcal{C}$ depends only on the sum of indices $K = m+n$, and we label consecutive adiabatic states in sector $\mathcal C$ by an integer $j$. At crossing $(m,n)$, the two adiabatic states participating in the avoided crossing have labels $j=m+n$ and $j=m+n+1$, which remain fixed along their respective adiabatic trajectories. At fixed $\tau$, neighboring crossings differ by $(\Delta m,\Delta n)=(1,1)$, so their index sums $m+n$ differ by two. Each crossing therefore accounts for two consecutive values of $j$. The occupation probabilities are obtained by evaluating the smooth C-sector profile at $\eta\approx \varepsilon j$, while the discrete sector sum counts every state separately. The total probability in each sector is given by the sum of the occupation probabilities over all sector indices:
\begin{align}
    \bar{P}_A &= \sum_{m\in\mathcal A}P_A(m), \quad
    \bar{P}_B = \sum_{n\in\mathcal B}P_B(n),\quad 
    \bar{P}_C = \sum_{j\in\mathcal C}P_C(j).
\end{align}
The sums are restricted to the disjoint asymptotic sectors, and $\bar{P}_A + \bar{P}_B + \bar{P}_C = 1$. The diagonal entropy is
\begin{equation}
\begin{aligned}
    S_d={}&-\sum_{m\in\mathcal A}P_A(m)\ln P_A(m)-\sum_{n\in\mathcal B}P_B(n)\ln P_B(n)\\
          &-\sum_{j\in\mathcal C}P_C(j)\ln P_C(j).
\end{aligned}
\label{eq:entropy_sector_sum}
\end{equation}
The profile maps suppress the constant coordinate offset $s_0=\ln(Z/\varepsilon)$. To evaluate them at the original state labels, replace $2\varepsilon m$, $2\varepsilon n$, $\varepsilon j$ inside every profile argument by $2\varepsilon m+s_0$, $2\varepsilon n+s_0$, $\varepsilon j+s_0$, respectively, before applying the displayed rescaling in Table~\ref{tab:explicit_profile_maps}. The same rule applies to the incoming profile definitions. With this in mind, we write the incoming probability profiles as
\begin{align}
    P_A^{\rm in}(m)&=f_A(2\varepsilon m),\\
    P_B^{\rm in}(n)&=f_B(2\varepsilon n),\\
    P_C^{\rm in}(j)&=f_C(\varepsilon j).
\label{eq:incoming_occupation_profiles}
\end{align}
Here $f_A,f_B,f_C$ are occupation profiles in the continuum description, expressed through the discrete sector indices. For the C sector, $j\simeq\eta/\varepsilon$
relates the discrete state label to the continuum coordinate. This approximation is used only in evaluating
the profile. The normalization sums over all individual adiabatic states. The incoming profiles are normalized by summing their occupations over all incident states. In the continuum approximation this gives
\begin{equation}
    \frac{1}{2\varepsilon}\int\mathrm{d}x
    \bigl[f_A(x)+f_B(x)+2f_C(x)\bigr]=1
\end{equation}
and absent incoming probability profiles are set to zero. 
Table \ref{tab:explicit_profile_maps} shows the incoming and outgoing profiles explicitly in terms of the incoming profiles. 
\subsection{Splitting of probability between two outgoing sectors}
\subsubsection{Regime \texorpdfstring{$0<\lambda<1$}{0 < lambda < 1}}
We first consider the case where a single sector loses states to two growing sectors and the initial distribution splits in two. We consider the case $0<\lambda<1$ which is one of the cases illustrated in Fig. \ref{fig:numerical_example}. Here, the incident $A$ wave is split into a reflected wave in the diabatic $B$ mode and a transmitted wave in the adiabatic sector $\mathcal{C}$ with equal $A$ and $B$ components. Integrating the outgoing profiles gives the total probabilities in each outgoing sector: 
\begin{align}
 \bar{P}_B^{\rm out}
 &=\int dn\,
 f_A\!\left(\frac{2\varepsilon(\lambda+1)}{\lambda-1}n\right)\nonumber\\
 &=\frac{1-\lambda}{1+\lambda}\int dm\,f_A(2\varepsilon m)
 =\frac{1-\lambda}{1+\lambda}\bar{P}_A^{\rm in},\label{P_B_1}\\[1mm]
 \bar{P}_C^{\rm out}
 &=\int dj\,
 f_A\!\left(\frac{\varepsilon(\lambda+1)}{\lambda}j\right)\nonumber\\
 &=\frac{2\lambda}{1+\lambda}\int dm\,f_A(2\varepsilon m)
 =\frac{2\lambda}{1+\lambda}\bar{P}_A^{\rm in},
  \label{eq:P_C_1}
 \end{align}
so the total probability $\bar{P}_{B}^{\rm(out)}+\bar{P}_{C}^{\rm(out)}=1$ is indeed conserved. Since $\bar P_A^{\mathrm{in}}=1$, the probability in each outgoing branch is independent of the shape of $f_A$.

Consider the case of N consecutive incoming A states with probability $1/N$ each, as is the case in the pendulum example. Here, $f_A(x) = 1/N$ on an interval of length $2\varepsilon N$ of the argument, and $f_A(x) = 0$ outside of that interval, \textit{i.e.} if the states $m_0,m_0+1,...,m_0+N-1$ are occupied initially, we have 
\begin{equation}
f_A(x)=
\begin{cases}
\dfrac{1}{N},
&2\varepsilon m_0\le x<2\varepsilon(m_0+N),\\[1mm]
0,&\text{otherwise}.
\end{cases}
\end{equation}
Since both outgoing profiles have unit coefficients, they retain height $1/N$. The integrated probabilities correspond to $N_B = \frac{1-\lambda}{1+\lambda}N$ occupied $\mathcal{B}$-states and $\frac{2\lambda}{1+\lambda}N$ occupied $\mathcal{C}$-states. The total occupied state count remains $N$, and every occupied state has probability $1/N$. Thus, the final diagonal entropy after the forward parameter sweep is the same as the initial value: $S_d^{\mathrm{f}}=S_d^\mathrm{in}=\ln N$. 
\subsubsection{Regime \texorpdfstring{$\lambda<-1$}{lambda < -1}}
In the case $\lambda < -1$, the adiabatic $\mathcal{C}$-sector loses states to both diabatic sectors analogously.  The total sector probabilities are
\begin{align}
 \bar{P}_A^{\rm out}&=
 \frac{\lambda+1}{2\lambda}\bar{P}_C^{\rm in},\\
 \bar{P}_B^{\rm out}&=
 \frac{\lambda-1}{2\lambda}\bar{P}_C^{\rm in}.
\label{eq:C_splitting_integrated_probabilities}
\end{align}
For $N$ equally occupied incoming C states, $f_C=1/N$ on an interval of length $\varepsilon N$ in the argument. The outgoing profiles
again retain height $1/N$, while the occupied A and B state counts are $N(\lambda+1)/(2\lambda)$ and
$N(\lambda-1)/(2\lambda)$, respectively. Their sum is $N$, so the
leading entropy change also vanishes in this splitting regime.
\subsection{Merging and entropic dilution of occupation probability}
\subsubsection{Regime \texorpdfstring{$-1<\lambda<0$}{-1 < lambda < 0}}
In this regime the diabatic sector $\mathcal{A}$ and adiabatic sector $\mathcal{C}$ lose states to the single growing diabatic sector $\mathcal{B}$, as in Fig.~\ref{fig:numerical_example} (b). 
Integrating the two contributions gives
\begin{equation}
 \begin{aligned}
 \bar{P}_B^{\rm out}\simeq{}&
 \int dn\,\frac{1+\lambda}{1-\lambda}
 f_A\!\left(\frac{2\varepsilon(1+\lambda)}{\lambda-1}n\right)\\
 &+\int dn\,\frac{2\lambda}{\lambda-1}
 f_C\!\left(\frac{2\varepsilon\lambda}{\lambda-1}n\right)\\
 ={}&\int dm\,f_A(2\varepsilon m)
    +\int dj\,f_C(\varepsilon j)\\
 ={}&\bar{P}_A^{\rm in}+\bar{P}_C^{\rm in}=1.
 \end{aligned}
 \label{eq:merging_into_B_integrated_probability}
\end{equation}
Each incoming contribution conserves its total probability: the
prefactor is compensated by the rescaling of the argument. The coefficients $(1+\lambda)/(1-\lambda)$ and $2\lambda/(\lambda-1)$
are positive and sum to one. The outgoing occupation
$P_B^{\rm out}(n)$ is therefore a weighted average of
the two incident occupation values appearing in the map. If the two incident occupation values contributing at
a given outgoing index are equal, the outgoing occupation
has that same value. If one incident sector is empty,
the outgoing occupation is reduced by the coefficient
multiplying the occupied input. The diagonal entropy change for two populated inputs thus depends on both incident profiles.

We now restrict the incoming ensemble to $\mathcal{A}$ by setting $f_C=0$, and assume again $N$ initially equally populated states, as is the case in Fig. \ref{fig:numerical_example}(b). Then, the outgoing profile is
\begin{equation}
 P_B^{\rm out}(n)=
 \frac{1+\lambda}{1-\lambda}
 f_A\!\left(\frac{2\varepsilon(1+\lambda)}{\lambda-1}n\right).
 \label{eq:A_to_B_profile}
\end{equation}
Its height, the probability per occupied B state, is
$(1+\lambda)/[(1-\lambda)N]$. Since
$\bar{P}_B^{\rm out}=1$, this means
$N_B =N(1-\lambda)/(1+\lambda)$ states in the outgoing sector are occupied. The change in diagonal entropy compared to the initial configuration is therefore
\begin{equation}
 \begin{aligned}
 \Delta S_d^{A\to B}&=
 \ln\!\left(\frac{1-\lambda}{1+\lambda}\right)>0.
 \end{aligned}
 \label{eq:A_to_B_entropy}
\end{equation}
For the forward sweep under protocol 2, we have $\lambda\approx- 0.440$ (see Table~\ref{tab:lattice_params} in Appendix~\ref{app:pendulum}) which gives $\Delta S_d \approx 0.946$ and relative change of diagonal entropy $\Delta S_d/\ln(30) \approx 0.278$. 

If we restrict instead to an ensemble of $N$ equally occupied C states, we set $f_A=0$ and use $f_C=1/N$ on an interval of length $\varepsilon N$ of the argument of that probability profile. The outgoing B profile is again uniform with height
$(-2\lambda)/[(1-\lambda)N]$, and again since $\bar{P}_B^\mathrm{out}=1$, the number of occupied states in the outgoing sector is $N(1-\lambda)/(-2\lambda)$, and the diagonal entropy change is
\begin{equation}
 \Delta S_d^{C\to B}=
 \ln\!\left(\frac{1-\lambda}{-2\lambda}\right)>0.
 \label{eq:C_to_B_entropy}
\end{equation}
\subsubsection{Regime \texorpdfstring{$\lambda>1$}{lambda > 1}}
For $\lambda>1$, $\mathcal{A}$ and $\mathcal{B}$ are incoming sectors and $\mathcal{C}$ is outgoing. Integrating the corresponding outgoing profile yields
\begin{equation}
 \bar{P}_C^{\rm out}=
 \bar{P}_A^{\rm in}+\bar{P}_B^{\rm in}=1.
 \label{eq:merging_into_C_integrated_probability}
\end{equation}
If we restrict to the incident $\mathcal{A}$ sector and again assume $N$ equally occupied states, the outgoing
occupation probability per state is $(\lambda+1)/(2\lambda N)$ over
$2\lambda N/(\lambda+1)$ C states. If only the B sector is incident, the probability is $(\lambda-1)/(2\lambda N)$ over $2\lambda N/(\lambda-1)$ outgoing C states.
Thus
\begin{equation}
 \begin{aligned}
 \Delta S_d^{A\to C}&=
 \ln\!\left(\frac{2\lambda}{\lambda+1}\right)>0,\\
 \Delta S_d^{B\to C}&=
 \ln\!\left(\frac{2\lambda}{\lambda-1}\right)>0.
 \end{aligned}
 \label{eq:merging_into_C_entropies}
\end{equation}
\section{Interpretation in terms of sector growth rates}
\begin{figure}[tbp]
    \centering
    \includegraphics[width=\linewidth]{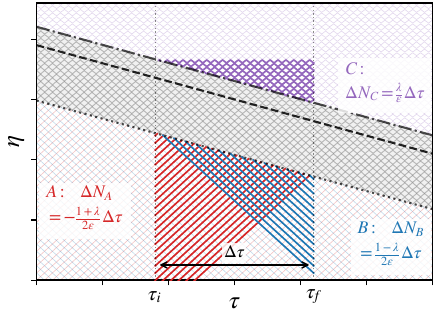}
    \caption{Illustration of the change of dimensionality of the two diabatic sectors below the separatrix zone and the adiabatic sector above the separatrix zone for a case with $0<\lambda<1$. The dashed line indicates the quantum separatrix contour for which $P = e^{-1}$, which serves as a reference within the finite separatrix zone; the dotted and dash-dotted lines show the crossings at an upper and lower probability threshold that bound the separatrix zone. The rising and falling diabats define the subspaces $\mathcal{A}$ (red) and $\mathcal{B}$ (blue) while $\mathcal{C}$ denotes the adiabatic sector above the quantum separatrix. The thick lines highlight states leaving or entering the respective sectors between $\tau_i$ and $\tau_f$, illustrating the decrease of $N_A$ and simultaneous growth of $N_B$ and $N_C$. In the continuum description, this yields the growth rates \eqref{dot_NA}-\eqref{dot_NC}.}
    \label{fig:state_space_dimensionality}
\end{figure}

\begin{table}
    \centering
    \begin{tabular}{c|c|c|c}
        \hline
         & $\dot N_A$ & $\dot N_B$ & $\dot N_C$ \\
        \hline
        $\lambda<-1$       & $+$ & $+$ & $-$ \\
        $-1<\lambda<0$     & $-$ & $+$ & $-$ \\
        $0<\lambda<1$      & $-$ & $+$ & $+$ \\
        $\lambda>1$        & $-$ & $-$ & $+$ \\
        \hline
    \end{tabular}
    \caption{Signs of the rates of change \eqref{dot_NA}-\eqref{dot_NC} of the sector dimensions
    $\mathcal{A}$, $\mathcal{B}$, $\mathcal{C}$ in the four regimes of $\lambda$.}
    \label{tab:growth_rates}
\end{table}
We denote by $\mathcal{A}, \mathcal{B}, \mathcal{C}$ the sectors of possibly populated diabatic states $\ket{E_m^A}$,$\ket{E_n^B}$ below the quantum separatrix and $\ket{E_j^C}$ adiabatic states above the separatrix. As we show in Appendix ~\ref{app:growth_rates} (in agreement with Ref. \cite{Stabel_2022}), the respective growth rates of the dimensions of the sectors are given by 
\begin{align}
    &\dot{N}_A = -\frac{(\lambda+1)}{2\varepsilon}\label{dot_NA}\\
    &\dot{N}_B = \frac{(1-\lambda)}{2\varepsilon}\label{dot_NB}\\
    &\dot{N}_C = \frac{2\lambda}{2\varepsilon}\label{dot_NC}
\end{align}
where $\sum_{k}\dot N_k=0$, $k\in\{A,B,C\}$ and $\dot{N}_k$  denotes the derivative with respect to $\tau$. We illustrate these rates in Fig.~\ref{fig:state_space_dimensionality} and Table~\ref{tab:growth_rates} shows which of the sectors are growing or shrinking depending on $\lambda$. In the case where one incident sector splits into two outgoing sectors, we can rewrite the total probability in each outgoing sector in terms of these growth rates of the sector dimensions. In the case $0<\lambda<1$, we can directly rewrite \eqref{P_B_1}, \eqref{eq:P_C_1} as 
\begin{align}
    &\bar{P}_B^\mathrm{out}=\frac{\dot N_B}{-\dot N_A}\\
    &\bar{P}_C^\mathrm{out}=\frac{\dot N_C}{-\dot N_A}
\end{align}
and, analogously, in the case $\lambda<-1$
\begin{align}
    &\bar{P}_A^\mathrm{out}=\frac{\dot N_A}{-\dot N_C} = \frac{\dot N_A}{\dot N_A+\dot{N}_B}\\
    &\bar{P}_B^\mathrm{out}=\frac{\dot N_B}{-\dot N_C}=\frac{\dot N_B}{\dot N_A+\dot{N}_B},
\end{align}
which agrees with the QKNH theorem \cite{Stabel_2022,qknh2}.

In the case where two incident sectors merge into one, the diagonal entropy does not necessarily increase. The increase can in fact vanish if we choose, for example, an initial configuration where both merging sectors have matching incident occupation values after the rescaling in Table~\ref{tab:explicit_profile_maps}. For general incident profiles, the entropy change depends on both profiles and their overlap in the outgoing sector. If we restrict to a single incident sector, the change follows a simple rule which we can again express in terms of the sector growth rates: 
\begin{equation}
    \Delta S^{\mathrm{in}\to\mathrm{out}}_d = -\ln C_{\mathrm{in}\to\mathrm{out}} = \ln\left|\frac{\dot{N}_{\mathrm{out}}}{\dot{N}_\mathrm{in}}\right|>0
\end{equation}
where $C_{\mathrm{in}\to\mathrm{out}}$ is the relevant transfer coefficient. Table \ref{tab:entropy_growth} shows the change in diagonal entropy depending on the incoming and outgoing sectors, according to this rule. In the entropy-producing regimes, two of the three connected state spaces shrink while the third one grows. Conservation of the total state-space dimension requires the growth of the single growing sector to compensate for the combined loss of the two shrinking sectors, and hence its growth rate necessarily exceeds the rate at which either incident sector individually shrinks. If we label the rate of change of the dimensions of the two shrinking sectors by $\dot N_{1,2}$, the rate of change of the single outgoing growing sector is $\dot N_{\mathrm{out}}=-(\dot N_1 + \dot N_2)$. An ensemble initially confined to one of the two shrinking sectors is therefore redistributed over a larger number of outgoing states, resulting in a dilution of the occupation probability.

\section{Summary and Discussion}
In this work, we have shown that the dynamics in the vicinity of a quantum separatrix can be described by a universal equation \eqref{eq:universaleq} in the limit of an infinitely dense network of avoided crossings. The equation is universal for quantum separatrices in the sense that it depends only on the parameter $\lambda$ which characterizes the relative growth of the connected dynamical sectors. The asymptotic scattering theory we derived here yields the total probability within these sectors in accordance with the QKNH theorem as well as the probability profiles within them. Hence, the diagonal entropy produced by crossing a quantum separatrix, in the case where we restrict to a single incoming sector, can be estimated by only the parameter $\lambda$. This entropy increase has a simple geometrical interpretation in terms of sector growth rates in this case. 
\begin{table}[t]
    \centering
    \begin{tabular}{c c c c}
        & Incident 
        & Outgoing
        & $\Delta S_d$ \\
        \hline
        $-1<\lambda<0$
        & $A$
        & $B$
        & $\displaystyle
          -\ln \dfrac{1+\lambda}{1-\lambda}
          =\ln\frac{\dot N_B}{-\dot N_A}$
          \\[3mm]

        $-1<\lambda<0$
        & $C$
        & $B$
        & $\displaystyle
          -\ln \dfrac{-2\lambda}{1-\lambda}
          =\ln\frac{\dot N_B}{-\dot N_C}$ \\[3mm]

        $\lambda>1$
        & $A$
        & $C$
        & $\displaystyle
          -\ln \dfrac{\lambda+1}{2\lambda}
          =\ln\frac{\dot N_C}{-\dot N_A}$ \\[3mm]

        $\lambda>1$
        & $B$
        & $C$
        & $\displaystyle
          -\ln \dfrac{\lambda-1}{2\lambda}
          =\ln\frac{\dot N_C}{-\dot N_B}$\\
    \end{tabular}
        \caption{Diagonal entropy change for an ensemble initially confined
    to a single shrinking state space. The entropy change is determined
    by the ratio of the growth rate of the outgoing state space to the
    magnitude of the growth rate of the incident state space.}
    \label{tab:entropy_growth}
\end{table}
As we showed in previous articles \cite{Stabel_2022,qknh2}, these sector growth rates directly relate to the growth rates of the corresponding classical phase space regions in the classical limit and their ratio corresponds well to the classical transition probability predicted by the KNH theorem \cite{Neishtadt_2017,henrard_capture_1982,LiouvilleControl} in this limit. 

The validity of the continuum theory beyond the extreme semiclassical limit is demonstrated in the pendulum examples where the avoided-crossing network is not extremely dense and the transition region is relatively narrow (as we show in \cite{qknh2}) but still the leading-order continuum theory predicts both the observed probability redistribution and the resulting entropy increase in good approximation. Rather than requiring an asymptotically dense continuum, the theory appears robust for more realistic systems containing only a finite number of participating avoided crossings. Such robustness is reminiscent of the well-known behavior of adiabatic approximations, whose asymptotic predictions often remain remarkably accurate far outside their formal domain of validity with exponentially small nonadiabatic corrections under appropriate conditions \cite{Molmer}. 
The present results complement previous studies of microscopic irreversibility associated with separatrix crossings. While earlier work established the emergence of irreversible probability redistribution and diagonal entropy growth in slowly driven isolated systems \cite{burkle_how_2019,buerkle_trimer,buerkle_qdimer1,buerkle_qdimer2}, the continuum theory developed here provides a local description of this redistribution in terms of effective scattering dynamics at the quantum separatrix. It links the choice of initially populated dynamical sector and geometry of the local crossing lattice to the irreversibility under forward-and-backward parameter sweeps. More generally, the emergence of a universal entropy law depending only on $\lambda$, if the system is initially restricted to one of the shrinking sectors, suggests that diagonal entropy production at quantum separatrices is a generic feature of slowly driven semiclassical systems, independent of their microscopic realization.

\section{Outlook}
A natural extension arises when the slowly varying control parameter itself becomes a dynamical coordinate, for example a reaction coordinate in molecular dynamics where the electronic Hamiltonian depends parametrically on the nuclear configuration while the nuclear coordinates evolve dynamically. How population is redistributed among electronic states then depends on pathways through non-adiabatic transition regions. Rather than making the system cross a quantum separatrix by sweeping an external control parameter, the evolving nuclear configuration itself could carry the system through a nonadiabatic transition region in the electronic spectrum with a dense network of avoided crossings. The effective local electronic dynamics within such transition regions could then potentially be replaced by the universal quantum separatrix scattering problem developed here which is characterized only by a parameter $\lambda$. In this sense, the quantum separatrix could play a role analogous to that of an elementary Landau-Zener transition, but for a dense network of avoided crossings. Exploring such an embedding of the effective quantum-separatrix scattering problem into coupled Born-Oppenheimer dynamics may provide a new route toward describing nonadiabatic transition regions in molecular dynamics. 

\emph{Data availability---}The main results of this work are analytical and are fully specified by the equations presented in the article. Details about the parameters and methods sufficient to reproduce the numerical examples are provided in the appendix. Further information is available from the authors upon reasonable request.
\appendix
\section{Relation of rectangular lattice coordinates and physical energy and time coordinates}
\label{app:lattice}
We consider two ladders of diabatic levels and linearize the semiclassical energy around a reference value $\sigma_0$ of the sweep parameter:
\begin{align}
\label{EA_semiclassical}
&E_m^A(\sigma)
\approx
E_0^A(\sigma_0)
+
(\sigma-\sigma_0)
\left.\frac{\partial E_0^A}{\partial\sigma}\right|_{\sigma_0}
+
m\Delta_A\\
\label{EB_semiclassical}
&E_n^B(\sigma)
\approx
E_0^B(\sigma_0)
+
(\sigma-\sigma_0)
\left.\frac{\partial E_0^B}{\partial\sigma}\right|_{\sigma_0}
+
n\Delta_B
\end{align}
where $\Delta_{A,B}$ are the level spacings $\hbar\Omega_{A,B}$ of the two diabatic ladders which are not, in general, equal. For a double well system, they could be for example the level spacings in the left and right well. For simplicity, we write $E^{\prime A}, E^{\prime B}$ for the derivative with respect to $\sigma$.

We assume $E^{\prime A}\ne E^{\prime B}$ and $\Delta_A+\Delta_B\ne 0$. The crossing labeled $(m,n)$ satisfies
$E_m^A(\sigma_{mn})=E_n^B(\sigma_{mn})$ which gives
\begin{equation}
\sigma_{mn}-\sigma_0
=
\frac{E_0^B-E_0^A+n\Delta_B-m\Delta_A}
     {E^{\prime A}-E^{\prime B}}.
\label{eq:app_sigma_mn}
\end{equation}
For the reference crossing $(0,0)$,
\begin{equation}
\sigma_{00}-\sigma_0
=
\frac{E_0^B-E_0^A}{E^{\prime A}-E^{\prime B}}.
\label{eq:app_sigma_00}
\end{equation}
Subtracting Eq.~\eqref{eq:app_sigma_00} from
Eq.~\eqref{eq:app_sigma_mn} removes the constant offset:
\begin{equation}
(E^{\prime A}-E^{\prime B})(\sigma_{mn}-\sigma_{00})
=
n\Delta_B-m\Delta_A.
\label{eq:app_sigma_index}
\end{equation}
\begin{figure}
    \centering
    \includegraphics[width=\linewidth]{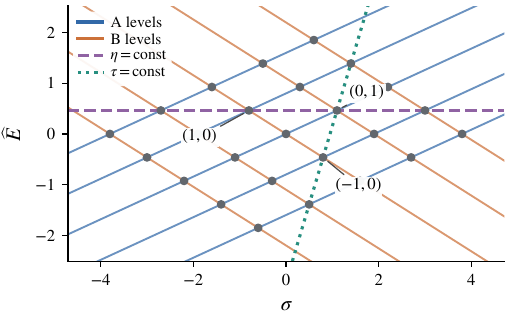}
    \caption{Crossing lattice in the $(\hat{E},\sigma)$ and $(\eta,\tau)$ coordinates. We show diabatic ladders $A$ (blue) and $B$ (orange), with unequal local level spacings, plotted against the parameter $\sigma$ using the shifted energy $\widehat E$. Additional dashed and dotted lines show illustrative lines of constant $\eta=\varepsilon(m+n)$ and constant $\tau=\varepsilon(n-m)$ respectively.}
    \label{fig:coordinate_map}
\end{figure}
We shift the energy coordinate by a common reference energy and a common parameter-dependent term,
\begin{equation}
\widetilde E
=
E-\frac{E_0^A+E_0^B}{2}
 -\frac{E^{\prime A}+E^{\prime B}}{2}
  (\sigma-\sigma_0).
\label{eq:app_energy_shift}
\end{equation}
The two shifted diabatic energies are
\begin{align}
\widetilde E_m^A
&=
\frac{E_0^A-E_0^B}{2}
+\frac{E^{\prime A}-E^{\prime B}}{2}
 (\sigma-\sigma_0)
+m\Delta_A,
\label{eq:app_shifted_A}\\
\widetilde E_n^B
&=
-\frac{E_0^A-E_0^B}{2}
-\frac{E^{\prime A}-E^{\prime B}}{2}
 (\sigma-\sigma_0)
+n\Delta_B.
\label{eq:app_shifted_B}
\end{align}
At their crossing these expressions are equal and the energy at the crossing is
\begin{equation}
\widetilde E_{mn}
=
\frac{m\Delta_A+n\Delta_B}{2}.
\label{eq:app_energy_index}
\end{equation}

We now introduce the continuum coordinates used in the main text,
\begin{equation}
\eta_{mn}=\varepsilon(m+n),
\qquad
\tau_{mn}=\varepsilon(n-m).
\label{eq:app_eta_tau}
\end{equation}
Their inverse is
\begin{equation}
m=\frac{\eta_{mn}-\tau_{mn}}{2\varepsilon},
\qquad
n=\frac{\eta_{mn}+\tau_{mn}}{2\varepsilon}.
\label{eq:app_inverse_indices}
\end{equation}
Substituting Eq.~\eqref{eq:app_inverse_indices} into
Eqs.~\eqref{eq:app_sigma_index} and
\eqref{eq:app_energy_index}, we obtain
\begin{align}
\sigma_{mn}-\sigma_{00}
&=
\frac{
(\Delta_B-\Delta_A)\eta_{mn}
+(\Delta_A+\Delta_B)\tau_{mn}}
{2\varepsilon(E^{\prime A}-E^{\prime B})},
\label{eq:app_sigma_eta_tau}\\
\widetilde E_{mn}
&=
\frac{
(\Delta_A+\Delta_B)\eta_{mn}
+(\Delta_B-\Delta_A)\tau_{mn}}
{4\varepsilon}.
\label{eq:app_energy_eta_tau}
\end{align}

These equations define a locally invertible map between the crossing coordinates and the physical parameter and shifted energy provided that both local level spacings
are nonzero and the two diabatic ladders have different slopes. Thus the $(\eta,\tau)$ coordinates provide a regular local description of the crossing lattice. In the general case $\Delta_A\neq\Delta_B$, each of the physical
coordinates $\sigma$ and $\tilde E$ depends on both $\eta$ and $\tau$.

An alternative shift makes the energy at each crossing proportional
to $\eta_{mn}$. In terms of the shifted energy defined in Eq. \eqref{eq:app_energy_shift},
we set
\begin{align}
\widehat E
&=
\widetilde E
-\frac{(\Delta_B-\Delta_A)(E^{\prime A}-E^{\prime B})}
       {2(\Delta_A+\Delta_B)}
 (\sigma-\sigma_{00})
\nonumber\\
&=
E-\frac{\Delta_B E_0^A+\Delta_A E_0^B}
        {\Delta_A+\Delta_B}
 -(\sigma-\sigma_0)
  \frac{\Delta_B E^{\prime A}+\Delta_A E^{\prime B}}
       {\Delta_A+\Delta_B}.
\label{eq:app_alternative_energy_shift}
\end{align}
At the crossing $(m,n)$, this gives
\begin{equation}
\widehat E_{mn}
=
\frac{\Delta_A\Delta_B}{\Delta_A+\Delta_B}(m+n)
=
\frac{\Delta_A\Delta_B}
     {\varepsilon(\Delta_A+\Delta_B)}\,\eta_{mn}.
\label{eq:app_alternative_energy_eta}
\end{equation}
Fig.~\ref{fig:coordinate_map} illustrates the coordinate mapping. 
\section{Transfer coefficients from matched asymptotics}
\label{sec:matched_asymptotics}
In the limit of small $|\nu|$ we need to consider three separate regimes of $s$ and use the method of matched asymptotics to find a global solution. The asymptotic scattering solution has the following form: 
\begin{widetext}
\begin{equation}
\label{eq:scattering_ansatz}
    \begin{tabular}{r c c c c}  & & $s\to-\infty$ & & $s \to +\infty$\\ 
    $\lambda < -1$\ :& & $T^A_\nu P_\nu \left(\begin{matrix} 1\\ 0\end{matrix}\right)e^{i\frac{\nu}{\lambda+1}s} +T^B_\nu P_\nu \left(\begin{matrix} 0\\ 1\end{matrix}\right)e^{i\frac{\nu}{\lambda-1}s}$ & &$P_\nu \left(\begin{matrix} 1\\ 1\end{matrix}\right)e^{i\frac{\nu}{\lambda}s}$\\
    $-1 < \lambda<0$\ :   & & $P_\nu \left(\begin{matrix} 1\\ 0\end{matrix}\right)e^{i\frac{\nu}{\lambda+1}s} + (R_\nu P_\nu + T_\nu P'_\nu)\left(\begin{matrix} 0\\ 1\end{matrix}\right)e^{i\frac{\nu}{\lambda-1}s}$ &  & $P'_\nu \left(\begin{matrix} 1\\ 1\end{matrix}\right)e^{i\frac{\nu}{\lambda}s}$ \\
    $0 < \lambda<1$\ : & & $P_\nu \left(\begin{matrix} 1\\ 0\end{matrix}\right)e^{i\frac{\nu}{\lambda+1}s} + R_\nu P_\nu\left(\begin{matrix} 0\\ 1\end{matrix}\right)e^{i\frac{\nu}{\lambda-1}s}$ & & $T_\nu P_\nu \left(\begin{matrix} 1\\ 1\end{matrix}\right)e^{i\frac{\nu}{\lambda}s}$ \\
    $\lambda > 1$\ : & & $P_\nu \left(\begin{matrix} 1\\ 0\end{matrix}\right)e^{i\frac{\nu}{\lambda+1}s} + P'_\nu \left(\begin{matrix} 0\\ 1\end{matrix}\right)e^{i\frac{\nu}{\lambda-1}s}$ & & $(T^A_\nu P_\nu + T^B_\nu P'_\nu) \left(\begin{matrix} 1\\ 1\end{matrix}\right)e^{i\frac{\nu}{\lambda}s}$ \;.
        \end{tabular}
\end{equation}
\end{widetext}
where $P_\nu, P'_\nu$ are independent inputs. We derive the reflection and transmission coefficients $R,T$ by the method of matched asymptotics. There is no single form of approximation that can exploit the smallness of $|\nu|$ accurately for all $s$. Instead we must consider three regions of $s$: $e^s\ll|\nu| \ \ (s\to-\infty)$ (region 1), $s\approx\ln|\nu|$ (region 2), and $e^{s} \gg |\nu| \ \ (s\to\infty)$ (region 3).

For sufficiently large negative $s$ such that $e^s \ll |\nu|$ (region 1), the second term on the right-hand side of (\ref{eq:effective_ODE}) can be treated as a perturbation to the first term. We find
\begin{align}
    \left(\begin{matrix}a(s)\\ b(s)\end{matrix}\right) =&  \left(\begin{matrix}Ae^{i\frac{\nu}{\lambda+1}s}\\ Be^{i\frac{\nu}{\lambda-1}s}\end{matrix}\right) \nonumber\\&- e^s \big(Ae^{i\frac{\nu}{\lambda+1}s}-Be^{i\frac{\nu}{\lambda-1}s}\big)\left(\begin{matrix}\frac{1}{1+\lambda}\\ \frac{1}{1-\lambda}\end{matrix}\right)+\mathcal{O}(\nu e^s, e^{2s})\;,
    \label{region1}
\end{align}
for arbitrary constants $A,B$. 

In the case $e^s \gg |\nu|$ (Region 3), on the other hand, we cannot simply treat the first term on the right of (\ref{eq:effective_ODE}) as a perturbation, because although the second matrix appears multiplied by the larger factor $e^s$, one of the matrix's eigenvalues is zero. To express $(a,b)^T$ in terms of the second matrix's two eigenvectors, we write
\begin{equation}
\label{region3_solution}
    \left(\begin{matrix}a(s)\\ b(s)\end{matrix}\right) = \frac{\alpha(s)}{2\lambda}\left(\begin{matrix}1\\ 1\end{matrix}\right) + \frac{\beta(s)}{2\lambda}\left(\begin{matrix}\frac{1}{1+\lambda}\\ \frac{1}{1-\lambda}\end{matrix}\right)\;.
\end{equation}
Equation (\ref{eq:effective_ODE}) is then equivalent to
\begin{align}
    \frac{d\alpha}{ds} &= \frac{i\nu}{\lambda}\Big(\alpha + \frac{\beta}{1-\lambda^2}\Big)\label{alphaeq}\\
    \frac{d\beta}{ds} - \frac{2\lambda e^s}{1-\lambda^2}\beta &= -\frac{i\nu}{\lambda}\Big(\alpha + \frac{1+\lambda^2}{1-\lambda^2}\beta\Big)\label{betaeq}\;.
\end{align}

Obtaining a solution to (\ref{alphaeq},\ref{betaeq}) to leading order in $\nu$ within all of region 3 requires multiple scale analysis, to distinguish terms which remain of order $\nu$ in size from terms which merely vary on the slow scale $\nu s$. We embed the line of $s$ in the plane $(\xi,\Xi)=(s,\nu s)$, and look for solutions $\alpha(\xi,\Xi)$ and $\beta(\xi,\Xi)$ to the coupled partial differential equations
\begin{align}\label{region1sol}
    \frac{\partial\alpha}{\partial \xi} &= -\nu \frac{\partial\alpha}{\partial \Xi} +\frac{i\nu}{\lambda}\Big(\alpha + \frac{\beta}{1-\lambda^2}\Big)\nonumber\\
    \frac{\partial\beta}{\partial \xi} - \frac{2\lambda e^\xi}{1-\lambda^2}\beta &=-\nu \frac{\partial\beta}{\partial \Xi} -\frac{i\nu}{\lambda}\Big(\alpha + \frac{1+\lambda^2}{1-\lambda^2}\beta\Big)\;.
\end{align}
The solutions only need to be evaluated on the line $(\xi,\Xi)=(s,\nu s)$, and on this line a term like $\nu\xi$, for example, is indistinguishable from $\Xi$. We use the freedom to select the larger number of integration constants in the two-dimensional equation to eliminate any terms like $\nu\xi$ from our solutions, and represent the same linear behavior as $\Xi$, so that it counts as order $\nu^0$ instead of order $\nu$, and is not neglected ---which it cannot be, because for large enough $s$ it will not be small. 
In the cases where $\lambda/(1-\lambda^2)<0$ and the second solution decays for $s\to+\infty$, we have
\begin{align}\label{region3_alpha}
    \alpha(s) &= E e^{i\frac{\nu}{\lambda}s} + i\nu F\frac{e^{-i\frac{\nu}{\lambda}\frac{1+\lambda^2}{1-\lambda^2}s}}  {\lambda(1-\lambda^2)}\mathrm{Ei}\big(\frac{2\lambda}{1-\lambda^2}e^s\big)+\mathcal{O}(\nu^2)\\\label{region3_beta}
    \beta(s) &= F e^{\frac{2\lambda}{1-\lambda^2}e^s} e^{-i\frac{\nu}{\lambda}\frac{1+\lambda^2}{1-\lambda^2}s} \nonumber\\&- \frac{i\nu}{\lambda}E e^{i\frac{\nu}{\lambda}s}e^{\frac{2\lambda}{1-\lambda^2}e^s}\mathrm{Ei}\big(-\frac{2\lambda}{1-\lambda^2}e^s\big) +\mathcal{O}(\nu^2)\;,
\end{align}
for arbitrary constants $E,F$ and $\mathrm{Ei}(x)$ is the exponential-integral function.  In the cases where $\lambda/(1-\lambda^2)>0$, the solution which grows for $s\to+\infty$ must be discarded, so we must force $F=0$.

In all ranges of $\lambda$, the terms involving the exponential-integral function decay rapidly for $s\to+\infty$, so that asymptotically we recover the above-separatrix solutions shown in the main text. We can then use the small-argument expansion
\begin{equation}
\label{Ei_expansion}
\mathrm{Ei}(x) = \gamma + \ln|x| + x +\mathcal{O}(x^2)
\end{equation}
where $\gamma = \lim_{n\to\infty}\sum_{k=1}^{n}\Big(\frac{1}{k}-\ln n\Big)\approx 0.577$ is the Euler-Mascheroni constant. We evaluate $\alpha(s)$ and $\beta(s)$ for $s= (1/2)\ln|\nu|<0$, and match to our Region 2 solution. We can solve (\ref{eq:effective_ODE}) in the intermediate Region 2 where $|e^s|\approx |\nu|$ starting from a zeroth-order solution in which $(a,b)^T$ is just constant. We obtain
\begin{align}\label{region2sol}
     \left(\begin{matrix}a(s)\\ b(s)\end{matrix}\right) =&  \left(\begin{matrix}C\\ D\end{matrix}\right) + i\nu s \left(\begin{matrix}\frac{C}{\lambda+1}\\ \frac{D}{\lambda-1}\end{matrix}\right) \nonumber\\&- (C-D) e^s \left(\begin{matrix}\frac{1}{1+\lambda}\\ \frac{1}{1-\lambda}\end{matrix}\right) +\mathcal{O}(\nu^2,\nu e^s,e^{2s})\;,
\end{align}
for constants $C,D$.

Comparing (\ref{region2sol}) with (\ref{region1}), we see that the two solutions smoothly match anywhere in the range around $s\sim (3/2)\ln|\nu|$ (where they are both leading terms in convergent series in $\nu$), as long as we identify
\begin{equation}
    (C,D) = (A,B)+\mathcal{O}(\nu^2)\;.
\end{equation}

Applying the relations
\begin{align}
    \alpha(s) &= (1+\lambda)a(s) - (1-\lambda)b(s)\nonumber\\
    \beta(s) &= (\lambda^2-1)[a(s)-b(s)]
\end{align}
in order to compare our solutions in regions 2 and 3, we can translate the region 2 solution (\ref{region2sol}) (after replacing $(C,D)$ with $(A,B)$) into
\begin{align}
    \alpha(s) &= (\lambda+1)A + (\lambda-1)B +i\nu s (A+B)\nonumber\\
    \beta(s) & = (A-B) \left(\lambda^2-1-{2\lambda e^s}+{i\nu \lambda s}\right) - (A+B){i\nu s}\;,
\end{align}
dropping terms of order $\nu^2$. We use the small argument expansion \eqref{Ei_expansion} of the exponential integral function and expand the coefficients in \eqref{region3_alpha},\eqref{region3_beta}, in orders of $\nu$, so $A = A_0+\nu A_1$ and analogously for $B,E,F$. Using a point in the matching region $s=1/2\ln(|\nu|)$ and sorting by orders of $\nu$, we find the following relation between $(A,B)$ and $(E,F)$: 
\begin{align}
    &E_0 = A_0\left(\lambda+1\right) + B_0\left(\lambda-1\right)\\
    &F_0 = (\lambda^2-1)(A_0 - B_0)\\
    &E_1 = (\lambda+1)A_1 + (\lambda-1)B_1 + i\frac{\Lambda(\lambda)}{\lambda}(A_0-B_0)\\
    &F_1 = (\lambda^2-1)(A_1-B_1) + \frac{i\Lambda(\lambda)}{\lambda}\left((\lambda+1)A_0+(\lambda-1)B_0\right)\\
    &\Lambda(\lambda) = \gamma + \ln\left|\frac{2\lambda}{1-\lambda^2}\right| 
\end{align}
Using the scattering ansatz \eqref{eq:scattering_ansatz} yields the leading order real transfer coefficients as well as imaginary $\mathcal{O}(\nu)$ corrections: 
\begin{equation}
\label{firstorder_TR}
    \begin{tabular}{c|c|c}
         $\lambda < -1$ & $T^A = 1 - i\nu\frac{\Lambda(\lambda)}{\lambda(\lambda+1)}$ & $T^B = 1 + i\nu\frac{\Lambda(\lambda)}{\lambda(\lambda-1)}$\\
         $-1<\lambda<0$&$R = \frac{1+\lambda}{1-\lambda}-2i\nu\frac{\Lambda(\lambda)}{(\lambda-1)^2}$ & $T= -\frac{2\lambda}{1-\lambda} +  2i\nu\frac{\Lambda(\lambda)}{(\lambda-1)^2}$\\
         $0<\lambda<1$&$R = 1 + 2i\nu\frac{\Lambda(\lambda)}{\lambda^2-1}$ & $T = 1 +i\nu\frac{\Lambda(\lambda)}{\lambda(\lambda+1)}$ \\
         $\lambda >1$ & $T^A=\frac{\lambda+1}{2\lambda} + i\nu\frac{\Lambda(\lambda)}{2\lambda^2}$ & $T^B = \frac{\lambda-1}{2\lambda} - i\nu\frac{\Lambda(\lambda)}{2\lambda^2}$
    \end{tabular}
\end{equation}
We transform back to the $(\eta,\tau)$ coordinates using 
\begin{align}
   &\frac{s}{\lambda\pm1}-\tau =\frac{\eta\mp\tau+s_0}{\lambda\pm1},\qquad \frac{s}{\lambda}-\tau=\frac{\eta+s_0}{\lambda}
\end{align}
where $s_0=\ln(Z/\varepsilon)$. In the following, particularly in eq. \eqref{scattering_eta}, the constant offset $s_0$ is suppressed in the equations since neglecting it leaves the total integrated sector probabilities and single-input entropy formulas unchanged. For merging of two ingoing profiles, the offset needs to be retained when mapping given occupations to the outgoing sector, so that the relative positions and overlap of the two contributions are evaluated correctly. To evaluate the following solution at the original coordinates, replace $\eta$ by  $\eta+s_0$ in every profile argument. Correspondingly, in Table~\ref{tab:explicit_profile_maps}, replace $2\varepsilon m,2\varepsilon n$ and $\varepsilon j$ by $2\varepsilon m+s_0$, $2\varepsilon n+s_0$ and $\varepsilon j + s_0$, respectively. With this in mind, the asymptotic form of the global solution to the effective probability transport equation is

\begin{widetext}
\begin{equation}
\label{scattering_eta}
    \begin{tabular}{r c c c c}  & & $s\to-\infty$ & & $s \to +\infty$\\
    \\
    $\lambda < -1$\ :& & $\left(\begin{matrix} T^A P\big(\frac{\lambda}{\lambda+1}(\eta-\tau)\big)\\ T^B P\big(\frac{\lambda}{\lambda-1}(\eta+\tau)\big)\end{matrix}\right)$ & & $P(\eta)\left(\begin{matrix} 1\\ 1\end{matrix}\right)$\\
    $-1 < \lambda<0$\ :   & & $\left(\begin{matrix} P(\eta-\tau)\\ R P\big(\frac{\lambda+1}{\lambda-1}(\eta+\tau)\big) + T P'\big(\frac{\lambda}{\lambda-1}(\eta+\tau)\big)\end{matrix}\right)$ &  & $P'(\eta) \left(\begin{matrix} 1\\ 1\end{matrix}\right)$ \\
    $0 < \lambda<1$\ : & & $\left(\begin{matrix} P(\eta-\tau)\\ RP\big(\frac{\lambda+1}{\lambda-1}(\eta+\tau)\big)\end{matrix}\right)$ & & $T P\big(\frac{\lambda+1}{\lambda}\eta\big) \left(\begin{matrix} 1\\ 1\end{matrix}\right)$ \\
    $\lambda > 1$\ : & & $\left(\begin{matrix} P(\eta-\tau)\\ P'(\eta+\tau)\end{matrix}\right)$ & & $\big(T^A P\big(\frac{\lambda+1}{\lambda}\eta\big) + T^B P'\big(\frac{\lambda-1}{\lambda}\eta\big)\big) \left(\begin{matrix} 1\\ 1\end{matrix}\right)$ \;.
        \end{tabular}
\end{equation}\end{widetext}
When integrating over $\eta$, total probability is of course conserved in all cases. For a term of the form $CP(\gamma\eta + \mu)$ in eq.~\eqref{scattering_eta}, the coefficient C changes the overall probability profile whereas $\gamma$ changes the length of the $\eta$-interval of occupied states: $\int d\eta\,  C P(\gamma\eta)=\frac{C}{|\gamma|}\int dx\,P(x)$. The total probability carried by an outgoing branch depends on both the scattering coefficient $R,T$ and the factors inside the argument of P. We use eq. \eqref{scattering_eta} to define the discrete-label sector transition maps in the main text.  

Extraction of the coefficients from numerical solutions of \eqref{eq:effective_ODE} shows that corrections to the leading order of the real part of the coefficients are of order $\nu^2$ and further corrections to the imaginary part are of order $\nu^3$ for all regimes of $\lambda$.
\begin{figure}
    \centering
    \includegraphics[width=1\linewidth]{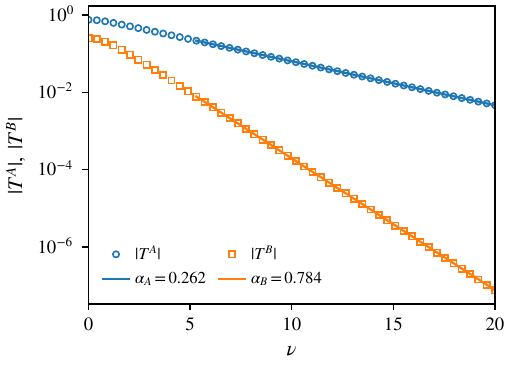}
    \caption{Numerically obtained transfer coefficients for $\lambda=2$ over a range of 50 equally spaced values of $\nu$. A fit in the interval $5<\nu\leq20$ reveals exponential decay rates for both transfer coefficients: $|T^j_\nu|\simeq C_j e^{-\alpha_j\nu}$ where $j\in\{A,B\}$, with fitted rates $\alpha_A$, $\alpha_B$ displayed in the legend.}
    \label{fig:large_nu_decay}
\end{figure}

In the limit of large $|\nu|$, on the other hand, the two eigenvalues of the combined matrix on the right side of eq. \eqref{eq:effective_ODE} in the main text differ by order $|\nu|\gg 1$ at all $s$ which suggests a semiclassical WKB approximation for $a_\nu(s),b_\nu(s)$. In this limit, the transfer coefficients are expected to become exponentially small as a function of $|\nu|$. Such suppression would smooth incident probability profiles by attenuating high-$\nu$ Fourier components. Numerical integration supports exponential suppression of high-frequency components, at least for the example $\lambda=2$ we show in Fig.~\ref{fig:large_nu_decay} where both transmission coefficients decay exponentially. Detailed analysis of this regime remains to be done. 

\section{Rates of change of the diabatic and adiabatic sectors}
\label{app:growth_rates}

We briefly review the geometrical origin of the growth rates of the
state spaces $\mathcal{A}$, $\mathcal{B}$, and $\mathcal{C}$ illustrated in Fig.~\ref{fig:state_space_dimensionality} which we originally derived in \cite{Stabel_2022}. The Landau-Zener probabilities on the (m,n) crossing lattice take the form of eq. \eqref{Pmn} which depends on
\begin{equation}
\tilde{s}=mX+nY+\ln(Z)=\eta+\lambda\tau+\ln(Z).
\end{equation}
Hence, lines of constant transition probability correspond to curves with $\tilde{s}=r_0=\mathrm{const.}$ and
\begin{equation}
    \eta_{r_0}(\tau)=r_0-\ln(Z)-\lambda\tau,
    \qquad
    \frac{d\eta_{r_0}}{d\tau}=-\lambda.
\end{equation}
In particular, the quantum separatrix contour $P=1/e$ corresponds to $r_0=0$. We may use this line as a reference contour
for counting states and write $\eta_{\rm sep}:=\eta_{r_0=0}$.

The diabatic $A$ and $B$ states correspond to fixed $m$ and $n$,
respectively. From the definitions of $\eta$ and $\tau$,
\begin{equation}
    \eta-\tau=2\varepsilon m,
    \qquad
    \eta+\tau=2\varepsilon n.
\end{equation}
Thus, an $A$ state moves along the characteristic
$\eta_A=\tau+\alpha$ with velocity $d\eta_A/d\tau=1$,
whereas a $B$ state moves along
$\eta_B=-\tau+\beta$ with velocity
$d\eta_B/d\tau=-1$ for constants $\alpha,\beta$.

The rates at which states cross the quantum separatrix are therefore
determined by the relative velocities of the diabatic characteristics
and the quantum separatrix. For the $A$ ladder this relative velocity is
\begin{equation}
    \frac{d}{d\tau}(\eta_A-\eta_{\mathrm{sep}})
    =1+\lambda,
\end{equation}
while for the $B$ ladder
\begin{equation}
    \frac{d}{d\tau}(\eta_{\mathrm{sep}}-\eta_B)
    =1-\lambda.
\end{equation}
Since neighboring states within each ladder are separated by
$\Delta\eta=2\varepsilon$ at fixed $\tau$, the corresponding state-counting rates are
\begin{equation}
    2\varepsilon\dot{N}_A=-(1+\lambda)
    \qquad
    2\varepsilon\dot{N}_B=1-\lambda.
\end{equation}
Conservation of the total state-space dimension,
$\dot N_A+\dot N_B+\dot N_C=0$, then yields
\begin{equation}
    2\varepsilon\dot{N}_C=2\lambda.
\end{equation}
Equivalently, with the definitions of $\lambda, \varepsilon$ in terms of the parameters $X$ and $Y$,
\begin{align}
\label{dot_NA_app}
    &\dot N_A=-\frac{2Y}{(X+Y)^2}\\\label{dot_NB_app}
    &\dot N_B=\frac{2X}{(X+Y)^2}\\\label{dot_NC_app}
    &\dot N_C=\frac{2(Y-X)}{(X+Y)^2}.
\end{align}
Although the crossover between the diabatic and adiabatic regimes has a finite width, the growth rates are independent of the particular constant-$s$ contour chosen for the state counting. All such contours are parallel and move with the same velocity $-\lambda$; changing $r_0$ therefore shifts the boundaries of the state spaces by a constant amount without changing their rates of change. To express these rates in terms of the physical sweep parameter $\sigma$, we use the local coordinate mapping in Eq.~\eqref{eq:app_sigma_eta_tau}. Along a reference contour
$\eta=r_0-\ln(Z)-\lambda\tau$, this gives
\begin{equation}
 \frac{d\sigma}{d\tau}
 =
 \frac{\Delta_A+\Delta_B-\lambda(\Delta_B-\Delta_A)}
      {2\varepsilon(E^{\prime A}-E^{\prime B})}
 =
 \frac{Y\Delta_A+X\Delta_B}
      {\varepsilon(X+Y)(E^{\prime A}-E^{\prime B})}.
\end{equation}
Applying the chain rule to Eqs.~\eqref{dot_NA_app}-\eqref{dot_NC_app}, with
$2\varepsilon=X+Y$, we obtain
\begin{equation}
 \frac{dN_A}{d\sigma}=-\Gamma Y,\qquad
 \frac{dN_B}{d\sigma}=\Gamma X,\qquad
 \frac{dN_C}{d\sigma}=\Gamma(Y-X),
\end{equation}
where
\begin{equation}
 \Gamma=
 \frac{E^{\prime A}-E^{\prime B}}
      {Y\Delta_A+X\Delta_B}
 =
 \frac{[\tilde S_A,\tilde S_B]_{E,\sigma}}
      {\pi\hbar\bigl(
       X\partial_E\tilde S_A+
       Y\partial_E\tilde S_B\bigr)}.
\end{equation}
The second equality follows from the semiclassical
relations $\Delta_j=\pi\hbar/\partial_E\tilde S_j$
and $E^{\prime }_j=-\partial_\sigma\tilde S_j/
\partial_E\tilde S_j$, with all quantities evaluated
at the reference crossing. These expressions reproduce
the subspace-dimensionality growth rates derived in
Ref.~\cite{Stabel_2022}, Eqs.~(B20)--(B24), upon identifying the sweep
parameter denoted by $\lambda$ in \cite{Stabel_2022} with $\sigma$ and its sector
labels $(A,C,B)$ with $(A,B,C)$ used here.

\section{Pendulum Model}
\label{app:pendulum}
\subsection{Classical Hamiltonian Model}

The classical Hamiltonian of the pendulum model we use for the numerical example in Fig. \ref{fig:numerical_example}  in the main text is given by

\begin{equation}
H(p,\phi, \alpha, \beta) = \frac{(p-\alpha(t))^2}{2\mu} - \beta^2(t)\cos\phi
\end{equation}

We obtain the corresponding quantum version by the usual quantization rules of replacing canonical variables with their corresponding Hermitian operators. Rotational orbit families of opposite orientation lie outside the libration region whose boundary is the separatrix. The energy of the separatrix contour corresponds to the height of the cosine barrier $E_s = \beta^2$.  Here, the parameter $\alpha$ translates the separatrix through the $(p,\phi)$ phase space, while $\beta$ determines the area enclosed by it. The sweep protocols are chosen such that the separatrix moves while simultaneously changing its enclosed area. 
\begin{figure}[t]
    \centering
    \includegraphics[width=\linewidth]{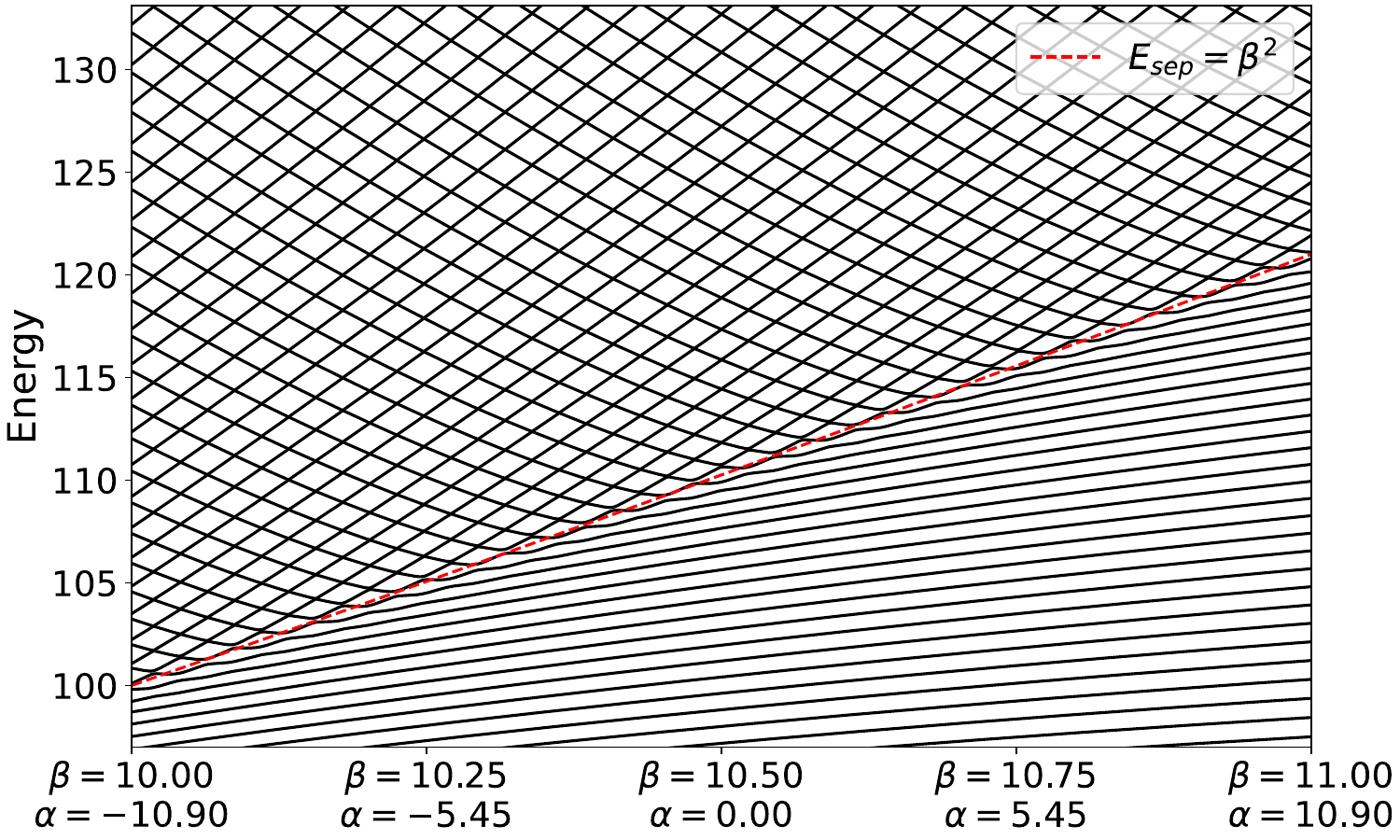 }
    \caption{Example of an instantaneous energy spectrum of the quantum pendulum system under a simultaneous $\alpha,\beta$ parameter sweep, showing avoided crossings above the classical separatrix energy. Here, we use $\mu=50$ and $\hbar=1$.  }
    \label{fig:spectrum}
\end{figure}
In the quantum version of this model, two sets of rotor states correspond to the energy-degenerate classical rotational orbits. Their degeneracy is lifted due to a coupling induced by above-barrier reflection. For energies above the classical separatrix energy, the spectrum shows many avoided crossings. Fig.~\ref{fig:spectrum} shows the spectrum for an illustrative example sweep protocol. 
\subsection{Parameter Sweep Protocols}
In the main text, we study the system's dynamics under two forward-and-backward parameter sweep protocols under which the parameters $\alpha(t)$ and $\beta(t)$ are varied slowly with a total sweep duration $T$ from initial values $(\alpha_i, \beta_i)$ at $t=0$ to final values $(\alpha_f,\beta_f)$ at the end of the forward sweep at $t=T/2$ and then return to their initial values at $t=T$. Although the Hamiltonian depends on the two parameters, we can parametrize their change in time by a single scalar control parameter $\sigma$: 
\begin{align}
\alpha(\sigma)
  &= \alpha_i + (\alpha_f-\alpha_i)\sigma,\\
\beta(\sigma)
  &= \beta_i + (\beta_f-\beta_i)\sigma.
\label{eq:pendulum_control_path}
\end{align}
The forward and reverse sweeps correspond to increasing $\sigma$ from zero to one and decreasing it back to zero, respectively:
\begin{equation}
\sigma(t)=
\begin{cases}
2t/T, & 0\leq t\leq T/2,\\
2-2t/T, & T/2\leq t\leq T.
\end{cases}
\label{eq:pendulum_sigma_cycle}
\end{equation}
Thus $\sigma(0)=\sigma(T)=0$ and $\sigma(T/2)=1$.

For the examples in Fig.~\ref{fig:numerical_example}, we use $\mu=50$, $\hbar=1$, $\alpha_i=-40.9$ and $\alpha_f=40.9$.
Protocol~1 has $\beta_i=10$ and $\beta_f=14$, so the area
enclosed by the classical separatrix grows during the forward
sweep and shrinks during the reverse sweep. Protocol~2
interchanges $\beta_i$ and $\beta_f$, reversing this order.

The forward sweep of protocol~1 realizes the regime
$0<\lambda<1$, in which an incident ensemble in sector $A$
splits between sectors $B$ and $C$. Protocol~2 realizes
$-1<\lambda<0$, in which sectors $A$ and $C$ shrink while
sector $B$ grows. Evaluation of the local lattice parameters
at the quantum separatrix (for more detail, see \cite{qknh2}) gives approximately $\lambda=\pm0.44$ for the forward sweep protocols~1 and~2, respectively.
\subsection{Numerical implementation of the quantum pendulum}
The Hamiltonian can be represented in the Fourier basis
\begin{equation}
    \varphi_n(\phi) = \braket{\phi|n} = \frac{e^{i n \phi}}{\sqrt{2\pi}}, \qquad n \in \mathbb{Z}.
\end{equation}
Using
\begin{equation}
    \hat{p} \ket{n} = \hbar n \ket{n},
\end{equation}
with the momentum operator $\hat{p}=-i\hbar\partial_\phi$ and
\begin{equation}
e^{\pm i\phi} \ket{n} = \ket{n \pm 1},
\end{equation}
the Hamiltonian can be written as
\begin{align}
\label{fourer_hamiltonian}
    \hat{H}(\sigma) &= \sum_{n=-\infty}^{\infty} \frac{(\hbar n - \alpha(\sigma))^2}{2\mu} \ket{n}\bra{n}\nonumber
    \\&- \frac{\beta^2(\sigma)}{2} \sum_{n=-\infty}^{\infty} \left( \ket{n}\bra{n+1} + \ket{n}\bra{n-1} \right).
\end{align}
For numerical implementation, we truncate the basis to $|n| \leq N$, which yields a finite $(2N+1)\times(2N+1)$ matrix representation of $\hat{H}$ and we set $\hbar=1$ if not stated otherwise. 

\begin{table}[tbp]
    \centering
    \begin{tabular}{c|c|c|c}
    \hline
    $\sigma$  & $X$ & $Y$ & $\lambda$\\
    \hline
    0.000 & 3.0015 & 1.1660 & -0.4404 \\
    0.250  & 3.0151&  1.1713  & -0.4404 \\
    0.500  & 3.0300& 1.1771 & -0.4404 \\
    0.750 & 3.0465& 1.1835  & -0.4404 \\
    1.000 & 3.0649& 1.1906  & -0.4404 \\
    \hline
    \end{tabular}
    \caption{Numerically extracted local parameters $X,Y$ and the resulting value of $\lambda$ for the quantum separatrix at five values of the sweep parameter $\sigma$ during the forward sweep under protocol 2.}
    \label{tab:lattice_params}
\end{table}
\subsection{Diagonal entropy and protocol averaging}
Let $\{|E_n(\sigma(t))\rangle\}$ denote the instantaneous energy
eigenstates of the Hamiltonian. For a coherently evolving
state $|\psi(t)\rangle$, the populations and diagonal entropy
are
\begin{align}
    p_n(t) &= |\langle E_n(\sigma(t))|\psi(t)\rangle|^2, \\
    S_d(t) &= -\sum_n p_n(t)\ln p_n(t),
\end{align}
$S_d$ coincides with the von Neumann entropy $S(t) = -\mathrm{Tr}(\bar{\rho}(t)\ln\bar{\rho}(t))$ of the effective density matrix obtained by discarding all off-diagonal elements of $\rho(t)=\ket{\psi(t)}\bra{\psi(t)}$ in the instantaneous energy basis. This discarding of off-diagonal elements could be achieved by time averaging the density matrix of the frozen system at each point in time yielding $\bar{\rho}(t)$, under the assumption of a nondegenerate spectrum. The von Neumann entropy of the full quantum state is of course preserved under unitary evolution. 
The diagonal entropy $S_{\mathrm{d}}(t)$ remains sensitive to coherent interference effects and may exhibit strong temporal fluctuations even in regimes where an effective probabilistic description would predict monotonic diagonal entropy growth. The diagonal entropy can even decrease through a concentration of amplitude.
\begin{figure}[t]
    \centering
    \includegraphics[width= \linewidth]{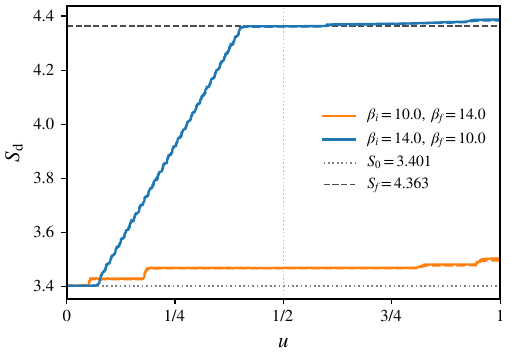}
    \caption{Diagonal entropy over normalized time $u=t/T_i$ for the two protocols shown in Fig. \ref{fig:numerical_example}. The solid lines show the diagonal entropy under protocol-averaged Schrödinger evolution; the colored dashed lines show the result from incoherent probabilistic evolution of the initial occupations. }
    \label{fig:S_diag}
\end{figure}
In our simulations, the initial state is an ensemble of $M$ equally weighted instantaneous energy eigenstates with indices in a range $\mathcal{I}$ (and $|\mathcal{I}|=M$), selected from the incident semiclassical ladder. Each ensemble member evolves independently under the same unitary dynamics with propagator $U(t,0)$, giving
\begin{align}
    \rho_{\mathrm{M}}(t)
    &= \frac{1}{M}\sum_{m\in\mathcal{I}}
       |\psi_m(t)\rangle\langle\psi_m(t)|, \\
    |\psi_m(t)\rangle
    &= U(t,0)|E_m(0)\rangle.
\end{align}
The corresponding populations are
\begin{equation}
    P_n^{\mathrm{M}}(t)
    = \frac{1}{M}\sum_{m\in\mathcal{I}}
      |\langle E_n(\sigma(t))|\psi_m(t)\rangle|^2.
\end{equation}
Their diagonal entropy is obtained by inserting
$P_n^{M}$ into the definition above, and the initial value is $S_d(0)=\ln M$. The mixture contains no coherence
between different initial ensemble members, but the
evolution of each member retains coherent interference.

In the pendulum examples in the main text, we additionally average over $R$ parameter sweep protocols with
different total durations $T_i$. All protocols follow the
same parameter path as a function of the normalized time
$u=t/T_i$, so the instantaneous Hamiltonian $H(\sigma(u))$ and its
instantaneous energy eigenbasis $\{|E_n(\sigma(u))\rangle\}$ are identical across realizations at
fixed $u$. Writing
\begin{equation}
    |\psi_m^{(i)}(u)\rangle
    = U_i(uT_i,0)|E_m(0)\rangle,
\end{equation}
the protocol-averaged populations of the equally weighted ensemble are
\begin{equation}
    \overline{P}_n^{M}(u)
    = \frac{1}{RM}
      \sum_{i=1}^{R}\sum_{m\in\mathcal{I}}
      |\langle E_n(\sigma(u))|\psi_m^{(i)}(u)\rangle|^2.
\end{equation}
We calculate the diagonal entropy of these averaged
populations,
\begin{equation}
    S_d^{\mathrm{p.a.}}(u)
    = -\sum_n
      \overline{P}_n^{M}(u)
      \ln \overline{P}_n^{M}(u).
\end{equation}
Protocol averaging suppresses interference by averaging the effectively random accumulated phases in the different protocols. For the examples considered here, protocol averaging of the quantum evolution of the ensembles reproduces the results one obtains with an incoherent effective probabilistic description. 
In the pendulum examples, we use an equally weighted ensemble of 30 instantaneous eigenstates of the initial Hamiltonian. We use the approximate matrix representation of the Hamiltonian \eqref{fourer_hamiltonian} with basis cutoff $N=300$. For protocol 1, we use the eigenstates $\ket{E_k}$ with numbers $k\in\{205,207,..,263\}$; for protocol 2 we use the eigenstates with numbers $k\in\{267,269,...,325\}$, and the index count starts at $0$. Here, every second state in the desired energy range is used since these states correspond asymptotically to diabats that encounter the quantum separatrix while the others correspond to diabats that evolve away from it. For both sweep protocols, we average the populations over 50 runs with sweep durations in the interval $[T_0-\delta T_0, T_0 + \delta T_0]$ for $T_0=10^4$ and $\delta = 0.01$.

\bibliography{bibliography}

\end{document}